\documentclass[journal,twocolumn]{IEEEtran}
\usepackage{amsmath}
\usepackage{diagbox}
\usepackage{graphicx}
\usepackage{float}
\usepackage{subfigure}
\newfloat{figtab}{htb}{fgtb}
\makeatletter
\newcommand\figcaption{\def\@captype{figure}\caption}
\newcommand\tabcaption{\def\@captype{table}\caption}
\makeatother

\usepackage{amssymb,amsthm}
\usepackage{color}
\usepackage{graphicx}
\usepackage{empheq}
\usepackage[linesnumbered,ruled,lined]{algorithm2e}
\usepackage[outdir=./]{epstopdf}
\usepackage{caption}
\usepackage{lipsum}
\usepackage{cite}
\usepackage{multirow}
\usepackage{fancyhdr}
\usepackage[colorlinks,
linkcolor=blue,       
anchorcolor=blue,  
citecolor=blue,        
]{hyperref}

\newtheorem{theorem}{Theorem}

\newtheorem{lemma}{Lemma}

\SetKwInput{Input}{input}
\SetKwInput{Output}{output}

\usepackage{array}
\newcolumntype{L}[1]{>{\raggedright\let\newline\\\arraybackslash\hspace{0pt}}m{#1}}
\newcolumntype{C}[1]{>{\centering\let\newline\\\arraybackslash\hspace{0pt}}m{#1}}
\newcolumntype{R}[1]{>{\raggedleft\let\newline\\\arraybackslash\hspace{0pt}}m{#1}}

\newtheorem{proposition}{Proposition}
\newtheorem{corollary}{Corollary}

\newtheorem{remark}{Remark}
\IEEEoverridecommandlockouts
\def\specialpapernotice#1{\if@confmode%
	\def\@specialpapernotice{{\sublargesize\textit{#1}\vspace*{1em}}}%
	\else%
	\def\@specialpapernotice{{\\*[1.5ex]\sublargesize\textit{#1}}\vspace*{-2ex}}%
	\fi}
\usepackage{amsmath,amsfonts}
\usepackage{algorithmic}
\usepackage{array}
\usepackage{textcomp}
\usepackage{stfloats}
\usepackage{url}
\usepackage{verbatim}
\usepackage{graphicx}
\usepackage{stfloats}
\usepackage{balance}
\usepackage{booktabs}
\usepackage{etoolbox}
\makeatletter\patchcmd{\@makecaption}{\scshape}{}{}{}

\usepackage{amsthm,amssymb,amsfonts}
\usepackage{bm}
\usepackage{cases}

\begin{document}
\title{Optimal Beamforming Structure and Efficient Optimization Algorithms for Generalized Multi-Group Multicast Beamforming Optimization}
\author{Tianyu Fang, Yijie Mao, \IEEEmembership{Member, IEEE}
\vspace{-0.8cm}
\thanks{This work has been supported in part by the National Nature Science Foundation of China under Grant 62201347; and in part by Shanghai Sailing Program under Grant 22YF1428400.
\par T. Fang and Y. Mao are with the School of Information Science and Technology, ShanghaiTech University, Shanghai 201210, China (e-mail:
\{fangty, maoyj\}@shanghaitech.edu.cn).}}


\maketitle

\begin{abstract}
In this work, we focus on solving non-smooth non-convex maximization problems in multi-group multicast transmission. Leveraging Karush-Kuhn-Tucker (KKT) optimality conditions and successive incumbent transcending (SIT) duality, we thoroughly analyze the optimal  beamforming structure for  a set of  optimization problems characterized by a general utility-based objective function.
By exploiting the identified optimal structure, we further unveil inherent low-dimensional beamforming structures within the problems, which are asymptotically optimal in various regimes of transmit signal-to-noise ratios (SNRs) or the number of transmit antennas. 
Building upon the discovered optimal and low-dimensional beamforming structures, we then propose highly efficient and toolbox-free optimization algorithms to solve a specific multi-group multicast optimization problem based on the weighted sum rate (WSR) utility function. 
The proposed algorithms first use the cyclic maximization (CM) framework to decompose the problem into multiple subproblems, each has an optimal or low-dimensional closed-form beamforming solution structure. Then, we propose the projected adaptive gradient descent (PAGD) algorithm  to compute the optimal Lagrangian dual variables for each subproblem.  
Numerical results show that the proposed algorithms maintain comparable or improved WSR performance compared to baseline  algorithms, while dramatically reducing the computational complexity. 
Notably, the proposed ultra-low-complexity algorithms based on low-dimensional beamforming structures achieve near optimal WSR performance with extremely low computational complexity. This complexity remains independent of the number of transmit antennas, making them promising and practical for extremely large multiple-input multiple-output (XL-MIMO) applications in 6G.
\end{abstract}


\section{Introduction}\label{Sec:intro}
\par 

Wireless communication systems are continually advancing to fulfill the escalating demands for high data rates, reliability, and energy efficiency. Within this landscape, the application of multicast strategies is crucial to effectively meet the communication requirements of numerous users seeking simultaneous access to identical data. Physical-layer multicast beamforming was first proposed in \cite{multicast2006} and has drawn much attention in recent years for its potential to support multi-group multicasting in various wireless services and applications, such as videoconferencing, mobile commerce, and intelligent transportation systems. With the emerging extremely large-scale multiple-input multiple-output (XL-MIMO) for 6G \cite{lu2023tutorial}, it is vital to develop low-complexity multi-group multicast beamforming solutions to address the high computational demands of these large-scale systems. 

\par The design of transmit beamforming in wireless communication systems aims to improve spectral or energy efficiency, typically involving the following two types of optimization problems: 1) total transmit power minimization under minimum  signal-to-interference-and-noise ratio (SINR) constraints for all user---\textit{the quality of service (QoS) problem}; 2) the system utility maximization subject to a total transmit power constraint---\textit{the utility problem}. These two types of problems are interconnected, and solutions to one can provide insight into the other. However, in multi-group multicast transmission, these two problems are generally NP-hard even in single-group scenarios \cite{multicast2006}. 
Various algorithms have been proposed to address multicast beamforming problems, including globally optimal and suboptimal algorithms. In \cite{lu2017efficient}, a globally optimal beamforming optimization algorithm employing the branch and bound (BB) method has been proposed for the QoS problem.
Although this global optimal algorithm exhibits attractive performance, it is tailored to a single-group scenario and is hindered by high computational complexity.
Therefore, state-of-the-art work focuses primarily on developing suboptimal algorithms capable of achieving near-optimal performance. 
\par 
Among the suboptimal algorithms for solving multicast beamforming problems, semi-definite relaxation (SDR) \cite{multicast2006,karipidis2008quality,chang2008approximation,xiang2012coordinated,christopoulos2014weighted} stands out as a popular convex relaxation (CR)-based method that achieves a near-optimal solution. However, as the size of the multicast network increases, the performance of SDR degrades quickly and the computational complexity increases sharply due to the auxiliary relaxation variables. To address these drawbacks, another category of optimization algorithms based on the convex approximation (CA) has emerged and received extensive investigation. Algorithms utilizing various techniques, including successive convex approximation (SCA), \cite{tran2013conic,christopoulos2015multicast,scutari2016parallel}, weighed minimum mean square error (WMMSE) \cite{yin2020rate,yin2022rate}, fractional programming (FP) \cite{Zhou2021,wang2021resource}, and majorization-minimization (MM) \cite{zhou2020intelligent,li2022secure}, have been proposed to solve various multi-group multicast beamforming optimization problems. Generally speaking, these CA-based algorithms are mathematically equivalent and share the same ability to find a near-optimal solution for the original problem \cite{zhang2023discerning}.
All the aforementioned algorithms either transform the original non-convex problem into a high-dimensional block-wise convex problem (i.e., WMMSE and FP) or construct a sequence of convex surrogate functions at a given point (i.e., SCA and MM). And each convex subproblem is solved using standard interior-point method (IPM), typically implemented by a dedicated  solver in optimization toolboxes, such as CVX \cite{grant2014cvx}.
However, the practical use of these algorithms is hampered by the undesirable computational complexity resulting from the iterative use of CVX optimization solvers.
\par 
To further reduce the computational complexity, several approaches shift towards closed-form and low-complexity beamforming designs for each convex subproblem obtained from the CA-based methods. Specifically, for the smooth QoS problems, alternating direction method of multipliers (ADMM) \cite{chen2017admm} and extragradient-based  \cite{Zhang2023} algorithms have been proposed to solve each subproblem. For the non-smooth utility problems, subgradient-based \cite{Mahmoodi2021} and log-sum-exp (LSE)-based \cite{zhou2020intelligent} algorithms have been introduced to handle each non-smooth subproblem. These methods generally offer lower complexity than the CVX-based algorithms. However, with the advent of XL-MIMO, which further boosts the number of transmit antennas by at least an order of magnitude compared to massive MIMO (e.g., several hundred or even thousands of transmit antennas), the computational complexity of these approaches still grows sharply with the number of transmit antennas. 
While certain low-complexity beamforming approaches, such as zero-forcing (ZF) \cite{sadeghi2017reducing} and weighted maximum ratio transmission (MRT) \cite{yu2018low}, are employed to reduce the  dimension  of the optimization problem, a unified analysis of the conditions under which these beamforming techniques are effective is currently lacking. 

\par 
The aforementioned ZF and MRT low-complexity beamforming approaches originate from the optimal beamforming structure \cite{Bjornson2014, Dong2020, fang2023optimal}. 
For unicast-only transmission,  the optimal beamforming structure was identified in \cite{Bjornson2014} for both QoS and general utility problems. It was demonstrated in \cite{Bjornson2014} that the optimal beamforming structures for these two problem types are equivalent. 
For multi-group  multicast transmission,  the optimal  beamforming structure was identified in \cite{Dong2020}, with a particular focus on the QoS and max-min fair (MMF) problems. Building upon the optimal multi-group multicast beamforming structure identified in \cite{Dong2020}, some ultra-low-complexity algorithms \cite{Zhang2022MMF,Shadi2022,Zhang2023} are proposed for large-scale communication networks. However, these algorithms mainly focus on  addressing the QoS problem or its inverse MMF problem by iteratively solving the QoS problem via a bisection search. This approach is not applicable for solving the utility problem due to the potentially high-dimensional search.
To the best of our knowledge, the optimal multi-group multicast beamforming structure for the generalized utility function-based maximization problems has not yet been identified. Additionally, efficient algorithms for directly solving these non-convex non-smooth problems with popular utility functions, such as weighted sum rate (WSR), geometric mean or harmonic mean of group rates are yet to be explored.

\par
This paper aims to bridge this gap by discovering the optimal multi-group multicast beamforming structure and efficient algorithms for solving the generalized utility function maximization problems. 
The key contributions of this paper are outlined as follows:
\begin{itemize}
	\item \textit{A thorough analysis of the optimal multi-group multicast beamforming structure}: By leveraging the Karush-Kuhn-Tucker (KKT) optimality conditions and the successive incumbent transcending (SIT) duality \cite{tuy2005robust}, we identify the optimal multi-group multicast beamforming structure for a generalized utility function-based maximization problem. In contrast to the approach in \cite{Dong2020}, our method first  derives out the optimal solution directly through the first-order optimality conditions of the original QoS problem. We then establish the SIT duality between the QoS problem and the utility problem, revealing the multi-group multicast beamforming for both types of problems shares the same optimal structure. 
	\item \textit{A comprehensive exploration of low-dimensional beamforming structures}: The revealed optimal beamforming structure provides valuable insights into the multi-group multicast beamforming design. This inspires us to  explore inherent low-dimensional beamforming structures that are asymptotically optimal in various regimes of transmit signal-to-noise (SNR) or the number of transmit antennas. These low-dimensional beamforming structures are particularly beneficial in reducing the computational complexity of beamforming design, especially in XL-MIMO systems.
	
     \item \textit{Development of highly efficient and optimization toolbox-free algorithms based on the identified beamforming structures}: Leveraging the identified optimal and low-complexity beamforming structures, we propose highly efficient and optimization toolbox-free algorithms to solve the non-smooth utility problem. Specifically, we first propose to utilize the cyclic maximization (CM) framework to decompose the problem into multiple subproblems, each has an optimal or low-dimensional closed-form beamforming solution structure.  Then, the projected adaptive gradient descent (PAGD) algorithm is proposed to compute the optimal Lagrangian dual variables for each subproblem.
     
     \item Numerical simulations demonstrate the superior efficiency of the proposed algorithms, as evidenced by their low central processing unit (CPU) time consumption, all while maintaining comparable or improved WSR performance compared to existing optimization algorithms. Surprisingly, the proposed ultra-low-complexity algorithms, leveraging low-dimensional beamforming structures, attain near-optimal WSR performance with remarkably low computational complexity. Importantly, this complexity remains independent of the number of transmit antennas, making them promising and practical for extensive applications in 6G, particularly in XL-MIMO scenarios.
\end{itemize}

\par \textit{Organization:} The rest of this paper is organized as follows. Section \ref{general optimal} and Section \ref{sec:lowDimBF} provide thorough analysis of the optimal and  low-complexity multi-group multicast beamforming structures, respectively, for the generalized utility problem. In Section \ref{Sec:WSR}, we leverage the CM framework and Lagrange duality to derive closed-form solution for each convex subproblem. Simulation results are presented in Section \ref{Sec:simulation}. In the end, Section \ref{Sec:concu} concludes the paper.

\par \textit{Notations:} Vectors and matrices are represented by bold lower-case and upper-case letters, respectively. The complex space is denoted by $\mathbb{C}$,  the real space by $ \mathbb{R} $, and $ \mathbb{R_+} $ denotes the set of real values larger than 0. Expectation over a random variable $ s $ is denoted as $ \mathbb{E}[s] $. The magnitude of a complex number $ x $ is expressed as $ |x| $. The circularly symmetric complex Gaussian distribution (CSCG) with zero mean and variance $ \sigma^2 $ is denoted as $ \mathcal{CN}(0,\sigma^2) $. The  conjugate transpose is represented by $ (\cdot)^H $. The optimal solution for a convex subproblem is denoted as $ (\cdot)^\star $. The local and global optimal solutions for the original non-convex problem are represented by $ (\cdot)^\diamond$ and $(\cdot)^\circ $, respectively. $ \mathrm{diag}\{ \mathbf y \} $ denotes a diagonal matrix with the entries of $ \mathbf y $ along the main diagonal, and $ \mathrm{blkdiag}\{\mathbf y_1,\cdots,\mathbf y_N\} $ is block-diagonal matrix with vector or matrices $\{\mathbf y_1,\cdots,\mathbf y_N\}$ along its main diagonal.

\section{Optimal Multi-group Multicast Beamforming Structure}\label{general optimal}

\subsection{System Model and Problem Formulation}
Consider a downlink multi-group multicast wireless communication network, where a base station (BS) equipped with $ L $ antennas simultaneously serving $ G $ non-overlapping user groups indexed by $ \mathcal G=\{1,\cdots,G\} $. In each user group $ g $, there are $ K_g $ single-antenna users indexed by $ \mathcal K_g=\{1,\cdots,K_g\},\forall g\in\mathcal G $. The total number of users in the system is  $ K=\sum_{g=1}^G K_g $. All users within the same group $ g $ require the same multicast stream $ s_g $. Without loss of generality, the transmit data stream vector $ \mathbf s=[s_1,\cdots,s_G]^T $ is assumed to have zero mean and identity variance, i.e., $ \mathbb{E}\{\mathbf s\mathbf s^H \}=\mathbf I_G $. Let $ \mathbf w_g\in\mathbb C^{L\times 1} $ be the corresponding beamforming vector for the stream $ s_g,\forall g\in\mathcal G $, the transmitted signal at the BS is $ \sum_{g=1}^{G}\mathbf w_g s_g $. The total transmit power is required to be less than or equal to the upperbound $ P_t $, i.e., $ \sum_{g=1}^G\|\mathbf w_g\|^2\leq P_t.  $     

\par Let $ \mathbf h_{gk}^H\in\mathbb{C}^{1\times L} $ be the channel vector from the BS to the user $ k $ in group $ g,\forall k\in\mathcal K_g, \forall g\in\mathcal G $, the signal received at user $k$  in group $ g $ is expressed as     
\begin{equation}
	\label{signal}
	y_{gk}=\mathbf h_{gk}^H\sum_{i=1}^{G} \mathbf w_i s_i+n_{gk},\forall k\in\mathcal K_g,\forall g\in\mathcal G,
\end{equation}
where $ n_{gk}\sim \mathcal{CN}(0,\sigma_{gk}^2) $ denotes the additive white Gaussian noise (AWGN)  at user $k$ in group $ g $ with $ \sigma_{gk}^2 $ denoting the noise power.

\par 
Each user in group $ g $ decodes the intended multicast stream $ s_g $ with the SINR of 
\begin{equation}
	\label{eq:SINRs}
	\gamma_{gk}=\left | \mathbf h_{gk}^H\mathbf w_g\right|^2\left(\sum_{i=1,i\neq g}^G\left |\mathbf h_{gk}^H\mathbf w_i\right|^2+  \sigma_{gk}^2\!\right)^{\!\!-1}\!\!\!\!\!\!,\forall k\in\mathcal K_g,\forall g\in\mathcal G.
\end{equation}
Consequently, the achievable rate for the multicast stream $ s_g $ is
\begin{equation}\label{rate}
	R_g=\min_{k\in\mathcal K_g} \left\{\log\left(1+\gamma_{gk}\right) \right\},\forall g\in\mathcal G.
\end{equation} 

\par In this study, our primary focus is on addressing a highly generalized multi-group multicast beamforming optimization problem, characterized by an utility function denoted as $ f(R_1,\cdots,R_G) $. Following \cite{Bjornson2014}, we assume that $ f(\cdot) $ is a continuous, strictly increasing function concerning the achievable SINRs of each user. It should be the operations that preserve concavity, such as non-negative weighted sums, pointwise minimum, and so on. The generalized utility  problem for multi-group multicast transmission is formulated as 
\begin{subequations}
	\label{P4}
	\begin{align}
		\mathcal U:	\max\limits_{\mathbf W}\,\, & f\left(R_1,\cdots,R_G\right)\\
	\label{P4:C1}	\text{s.t.}\,\,& \mathrm{Tr}\left(\mathbf W\mathbf W^H\right)\leq P_t,
	\end{align}
\end{subequations}
where $ \mathbf W\triangleq[\mathbf w_1,\cdots,\mathbf w_G] $. This problem poses two major challenges: the presence of multiple non-convex fractional SINR expressions \eqref{eq:SINRs} and the non-smooth nature of the achievable rate expressions \eqref{rate} for multicast streams. Notably, the multicast beamforming design is inherently NP-hard, even when the problem \eqref{P4} is reduced to the single-group setting where inter-group interference is absent \cite{multicast2006}.

\par  To address the non-smooth characteristics of problem \eqref{P4}, we first investigate the  more tractable QoS problem in \eqref{P2}, and subsequently extend the findings to problem \eqref{P4} by exploring their interrelations. The QoS problem is formulated as follows\begin{subequations}
	\label{P2}
	\begin{align}
	\mathcal P:	\min\limits_{\mathbf W}\,\, & \mathrm{Tr}\left(\mathbf W\mathbf W^H\right)\\
\label{P2:C1}		\text{s.t.}\,\,& \gamma_{gk}\geq \alpha_{g}, \forall k\in\mathcal K_g,\forall g\in\mathcal G, 
\end{align}
\end{subequations}
where $ \alpha_g $ refers to the QoS threshold for multicast stream $ s_g $. Users within the same group share a common QoS threshold since the achievable rate is constrained by the worst-case user within each user group. This differs from the unicast-only scenario, where each user has an individual QoS threshold. Problem \eqref{P2} has been proven to be non-convex and NP-hard \cite{karipidis2008quality}.
\par
In the next subsections, we focus on identifying the optimal multicast beamforming structure for both problem \eqref{P4} and problem \eqref{P2}, using an approach different from the iterative SCA method in \cite{Dong2020}.

\subsection{Optimal Multicast Beamforming Structure for QoS problem}
 
In this subsection, we identify the optimal multicast beamforming structure for the power minimization problem \eqref{P2}. Specifically, we begin by using Lemma \ref{lemma LICQ} to show that the KKT conditions of (\ref{P2}) are necessary conditions for all stationary points. Following this, we directly unveil the optimal beamforming structure of \eqref{P2} based on its KKT conditions.

\begin{lemma}\label{lemma LICQ}
	Linear independence constraint qualification (LICQ) \cite{peterson1973review} holds for problem \eqref{P2} when all channel vectors $ \{\mathbf h_{gk},\forall k\in\mathcal K_g,\forall g\in\mathcal G\} $ are linearly independent. 
\end{lemma}

\textit{Proof:} Constraint \eqref{P2:C1} can be rewritten as
\begin{equation}
	\label{Rewrite SINR}\underbrace{\sum_{i=1,i\neq g}^G\frac{1}{\sigma_{gk}^2}\left|\mathbf h_{gk}^H\mathbf w_i\right|^2+1-\frac{1}{\alpha_{g}\sigma_{gk}^2}\left|\mathbf h_{gk}^H\mathbf w_g\right|^2}_{c_{gk}(\mathbf W)}\leq 0.
\end{equation}  
Denote the left-hand side of \eqref{Rewrite SINR} as $ c_{gk}(\mathbf W) $, then the gradient of $ c_{gk}(\mathbf W) $ with respect to $ \mathbf W $ is given as 
\begin{equation*}\label{key}
	\nabla c_{gk}(\mathbf W)=\frac{2}{\sigma_{gk}^2}\mathbf h_{gk}\mathbf h_{gk}^H\left[\mathbf w_1,\cdots,\frac{-1}{\alpha_{g}}\mathbf w_g,\cdots, \mathbf w_G \right].
\end{equation*}
All gradients $ \{\nabla c_{gk}(\mathbf W) ,\forall k\in\mathcal K_g,\forall g\in\mathcal G\} $ are linearly independent and therefore satisfy LICQ if the channel vectors $ \{\mathbf h_{gk},\forall k\in\mathcal K_g,\forall g\in\mathcal G\} $ exhibit linear independence. $ \hfill\blacksquare $

\begin{remark}\label{channel condituions}
	The power minimization problem \eqref{P2} for multi-group multicasting beamforming would be infeasible when the channel vectors from different groups are linearly dependent, since the interference from other groups can not be eliminated by the beamforming vectors. Without loss of generality, we assume that all channel vectors $ \{\mathbf h_{gk}\} $ are linearly independent, and therefore problem \eqref{P2} is feasible. This assumption holds for widely adopted channel models, such as Rayleigh fading. 
\end{remark}
 
Lemma \ref{lemma LICQ} implies that the KKT conditions are necessary conditions for any stationary point for problem \eqref{P2}. Therefore, we are able to analyze the problem \eqref{P2} using the KKT optimality conditions. By introducing a set of Lagrange multipliers $\{ \lambda_{gk}\} $ for the corresponding reformulated SINR constraints \eqref{Rewrite SINR}, we define the Lagrangian function of \eqref{P2} as
 \begin{equation}\label{LP2}
 	\begin{split}
 	&	\mathcal{L}_{\eqref{P2}}(\mathbf W,\bm \lambda)=\frac{1}{2}\sum_{g=1}^G \|\mathbf w_G\|^2\\
 		&\!+\!\frac{1}{2}\!\sum_{g=1}^G\sum_{k=1}^{K_g}\lambda_{gk}\!\! \left(\sum_{i=1,i\neq g}^G\!\frac{1}{\sigma_{gk}^2}|\mathbf h_{gk}^H\mathbf w_i|^2+\!1\!-\frac{1}{\alpha_{g}\sigma_{gk}^2}|\mathbf h_{gk}\mathbf w_g|^2\!\right)\!,\notag
 	\end{split}
 \end{equation}
 where $ \bm\lambda\triangleq[\bm\lambda_1,\cdots,\bm\lambda_G] $ with $ \bm\lambda_g=[\lambda_{g1},\cdots,\lambda_{gK_g}]^T $. The first-order derivative of $ \mathcal{L}_{\eqref{P2}}(\mathbf W,\bm \lambda) $ with respective to $ \mathbf w_g $ is given by
 \begin{equation}\label{der}
 	\frac{\partial \mathcal L_{\eqref{P2}}}{\partial \mathbf w_g}=\mathbf w_g+\sum_{i\neq g}^G\sum_{k=1}^{K_i}\frac{\lambda_{ik}}{\sigma_{ik}^2}\mathbf h_{ik}\mathbf h_{ik}^H\mathbf w_g-\sum_{k=1}^{K_g}\frac{\lambda_{gk}}{\alpha_{g}\sigma_{gk}^2}\mathbf h_{gk}\mathbf h_{gk}^H\mathbf w_g.
 \end{equation}
 Therefore, for any stationary point $ \mathbf W^\diamond $ of problem \eqref{P2}, there exists a set of Lagrange multipliers $ \{\lambda_{gk}^\diamond\} $ satisfying the stationary conditions, i.e., $ \partial \mathcal L_{\eqref{P2}}/\partial \mathbf w_g^\diamond=\mathbf 0 $, which leads to 
\begin{equation}\label{KKT}
	\left(\!\mathbf I_L \!+\!\sum_{i=1}^G\!\sum_{k=1}^{K_i}\frac{\lambda_{ik}^\diamond}{\sigma_{ik}^2}\mathbf h_{ik}\mathbf h_{ik}^H\!\right)\!\mathbf w_g^\diamond\!=\!\sum_{k=1}^{K_g}\frac{\lambda_{gk}^\diamond}{\sigma_{gk}^2}\left(1+\frac{1}{\alpha_{g}}\right)\mathbf h_{gk}\mathbf h_{gk}^H\mathbf w_g^\diamond.
\end{equation}
Equation \eqref{KKT} is obtained from \eqref{der} by adding and subtracting the term $ \sum_{k=1}^{K_g}\frac{\lambda_{gk}^\diamond}{\sigma_{gk}^2}\mathbf h_{ik}\mathbf h_{ik}^H\mathbf w_g^\diamond $, and then setting it to zero. The derived locally optimal beamforming solution is given as
\begin{equation}\label{locally}
	\mathbf w_g^\diamond\!=\!\left(\!\mathbf I_L \!+\!\!\sum_{i=1}^G\sum_{k=1}^{K_i}\!\frac{\lambda_{ik}^\diamond}{\sigma_{ik}^2}\mathbf h_{ik}\mathbf h_{ik}^H\!\right)^{\!\!-1}\!\sum_{k=1}^{K_g}\!\frac{\lambda_{gk}^\diamond}{\sigma_{gk}^2}\!\left(\!1\!+\!\frac{1}{\alpha_{g}}\!\right)\mathbf h_{gk}\mathbf h_{gk}^H\mathbf w_g^\diamond.
\end{equation}
The global-optimal beamforming solution of problem \eqref{P2} aligns with the beamforming structure in \eqref{locally}, as it belongs to one of the local-optimal solutions of \eqref{P2}. For simplicity, we write out the optimal beamforming structure in matrix form as shown in the following Theorem \ref{theorem global}.
\begin{theorem}\label{theorem global}
		The optimal multi-group multicast beamforming structure for problem \eqref{P2} is
	\begin{equation}\label{theorem optimal structure}
		\mathbf w_g^\circ=\left( \mathbf I_L+\sum_{i=1}^G\mathbf H_i \bm\Theta_i^\circ \mathbf H_i^H  \right)^{-1} \mathbf H_g\mathbf d_g^\circ, \forall g\in\mathcal G,
	\end{equation} 
where $ \mathbf H_i\triangleq [\mathbf h_{i1},\cdots,\mathbf h_{iK_i}]$, $ \bm\Theta_i^\circ\triangleq \mathrm{diag}\{ \theta_{i1}^\circ,\cdots, \theta_{iK_i}^\circ\} $ with $\theta_{ik}^\circ=\frac{\lambda_{ik}^\circ}{\sigma_{ik}^2}$, $ \mathbf d_g^\circ\triangleq [d_{g1}^\circ,\cdots,d_{gK_g}^\circ]^T $ with $ d_{gk}^\circ=\frac{\lambda_{gk}^\circ}{\sigma_{gk}^2}\left(\!1\!+\!\frac{1}{\alpha_{g}}\!\right)\mathbf h_{gk}^H\mathbf w_g^\circ$, and $ \{\lambda_{gk}^\circ\} $ are corresponding optimal dual variables. 

\end{theorem}
 Theorem \ref{theorem global} coincides with the optimal structure discovered in \cite{Dong2020}. But the proof is much simpler and more intuitive, since it is built on the LICQ and KKT conditions of the original problem.
 \par  By directly setting equation \eqref{der} equal to zero, the optimal beamfroming structure for problem \eqref{P2} has another equivalent form, as given in the following Corollary \ref{optimal structure}.  
\begin{corollary} \label{optimal structure}
	The optimal multicast beamforming solution $ \mathbf w_g^\circ $ in \eqref{theorem optimal structure} has the following equivalent form
	\begin{equation}\label{equivalent form}
		\mathbf w_g^\circ=\left( \mathbf I_L+\sum_{i=1,i\neq g}^G\mathbf H_i \bm\Theta_i^\circ \mathbf H_i^H  \right)^{-1} \mathbf H_g\tilde{\mathbf d}_g^\circ, \forall g\in\mathcal G,
	\end{equation}
where $  \bm\Theta_i^\circ $ is the same as in \eqref{theorem optimal structure}, and $ \tilde{\mathbf d}_g^\circ\triangleq [\tilde{d}_{g1}^\circ,\cdots,\tilde{d}_{gK_g}^\circ]^T $ with $ \tilde{d}_{gk}^\circ=\frac{\lambda_{gk}^\circ}{\alpha_{g}\sigma_{gk}^2}\mathbf h_{gk}^H\mathbf w_g^\circ $.
\end{corollary}
\par
Although the parameters $ \bm\Theta_g^\circ $ and $ \mathbf d_g^\circ $ are challenging to obtain due to the NP-hard nature of problem \eqref{P2}, the optimal multicast beamforming structure brings valuable insights to the beamforming design. This is particularly evident when it is used to reduce the dimensions of optimization variables, which we will discuss later. 
\par
So far, we have attained the optimal beamforming structure for the power minimization problem \eqref{P2}. However, it remains challenging to solve the general utility problem \eqref{P4}. In the following subsections, we will identify the optimal multicast beamforming structure for problem \eqref{P4} and discuss some valuable insights. This is a major contribution of this work.

\subsection{Optimal Multi-group Multicast Beamforming Structure for General Utility Function Maximization}
In this subsection, we aim to identify the optimal beamforming structure for the generalized utility problem \eqref{P4} using the SIT duality approach. Introduced in \cite{tuy2005robust}, SIT duality is an optimization approach for calculating the global-optimal solution for non-convex optimization problems. It has shown its effectiveness in solving various resource allocation problems, as demonstrated in \cite{Boh2019,Matth2022Globally}. To illustrate the SIT principle, we first exchange the objective function and the constraint in \eqref{P4}, resulting in the following SIT dual problem
\begin{subequations}
	\label{P5}
	\begin{align}
		\min\limits_{\mathbf W}\,\, &\mathrm{Tr}\left(\mathbf W\mathbf W^H\right) \\
		\text{s.t.}\,\,& f\left(R_1,\cdots,R_G\right)\geq f\left(\beta_1^\circ,\cdots,\beta_G^\circ\right),
	\end{align}
\end{subequations}
where $ \beta_g^\circ,\forall g\in\mathcal G $ denotes the optimal achievable rate for the multicast stream $ s_g $ at the global-optimal solution of the original problem \eqref{P4}. Given the assumption that $ f(R_1,\cdots,R_G) $ is strictly increasing with respect to $ \{R_1,\cdots,R_G\} $, problem \eqref{P5} can be further reformulated as problem \eqref{P2} where the QoS threshold is given as $ \alpha_{g}^\circ=\exp(\beta_{g}^\circ)-1 $. The SIT principle tells that the optimal solution of \eqref{P4} can be obtained by solving a sequence of power minimization problems \eqref{P2} with increasing SINR constraints $ \alpha_{g}^\circ $. The optimal $ \alpha_g^\circ $ can be obtained by a $ G- $dimensional bisection search.

 Based on the principle of SIT, the SIT duality between problem \eqref{P4} and problem \eqref{P2} can be established. Specifically, let $ \mathbf H=[\mathbf H_1,\cdots,\mathbf H_G] $ and $ \bm\sigma=[\sigma_{11},\cdots,\sigma_{GK_G}]^T, $ a mapping of the general utility problem \eqref{P4} is defined as
 \begin{equation}\label{key}
 	  \mathcal U: \mathbb{R_+} \rightarrow \mathbb R_+^{G},\,\,\bm\beta^\circ=\mathcal U(P_t| \mathbf H,\bm\sigma),
 \end{equation}
 where $ \bm\beta^\circ=[\beta_{1}^\circ,\cdots,\beta_{G}^\circ]^T $. $\mathcal U(P_t|\mathbf H,\bm\sigma)$ solves problem \eqref{P4} based on the input parameter $P_t$, the output corresponds to using the optimal solution $ \mathbf W^\circ $  to compute the optimal rate vector $ \bm\beta^\circ$.   Also, define the corresponding mapping of the power minimization problem \eqref{P2} as 
 \begin{equation}\label{key}
 	\mathcal P: \mathbb{R}_+^G \rightarrow \mathbb R_+,\,\,P_t=\mathcal P(\bm\beta^\circ|\mathbf H,\bm\sigma).
 \end{equation} Similarly, $\mathcal P(\bm\beta^\circ|\mathbf H,\bm\sigma)$ solves problem \eqref{P2} based on the input parameters $\bm\beta^\circ$, the output corresponds to the minimized transmit power at the  optimal solution. Then, the SIT dual relation between problem \eqref{P4} and problem \eqref{P2} is described in the following Proposition \ref{claim SIT dual}.
\begin{proposition}\label{claim SIT dual}
	The SIT duality between problem \eqref{P4} and problem \eqref{P2} is established as
	\begin{equation}\label{SIT dual}
	\begin{split}
		P_t&=\mathcal P(\mathcal U(P_t\left|\mathbf H,\bm\sigma\right ) \left|\mathbf H,\bm\sigma\right )\\
			\bm\beta^\circ&=\mathcal U(\mathcal P(\bm\beta^\circ\left| \mathbf H,\bm\sigma\right )\left|\mathbf H,\bm\sigma\right)
	\end{split}
	\end{equation}
\end{proposition}
\textit{Proof:} This conclusion can be obtained directly by using the proofs of SIT duality in existing works \cite{tuy2005robust,Boh2019,Matth2022Globally}. 
Due to space limitations, we omit the proof details in this work. $ \hfill\blacksquare $

The SIT duality \eqref{SIT dual} implies that the general utility problem \eqref{P4} can be solved by searching over rate targets $ \bm\beta $, such that the optimal objective value of solving \eqref{P2} for a given rate target $ \bm\beta^\circ $ is equivalent to the constraint upperbound $ P_t $ in \eqref{P4:C1}. Therefore, problems
\eqref{P4} and \eqref{P2} share the same optimal beamforming structure as shown in Theorem \ref{theorem general}.
\begin{theorem}\label{theorem general}
	The optimal beamforming solution structures for both problem \eqref{P4} and problem \eqref{P2} are equivalent to
	\begin{equation}\label{general structure}
			\mathbf W^\circ=\left( \mathbf I_L+\mathbf H \bm\Theta^\circ \mathbf H^H  \right)^{-1} \mathbf H\mathbf D^\circ
	\end{equation}
where $ \bm\Theta^\circ=\mathrm{blkdiag}\{\bm\Theta_1^\circ,\cdots,\bm\Theta_G^\circ\}  $ and $ \mathbf D^\circ=\mathrm{blkdiag}\{\mathbf d_1^\circ,\cdots,\mathbf d_G^\circ\} $ are some  parameters.
\end{theorem}
Although it remains challenging to determine the optimal achievable rate target $ \bm\beta^\circ $, the SIT duality helps to identify the optimal beamforming structure for  problem \eqref{P4}.

\subsection{Insights from the Optimal Beamforming Structure}
To better characterize the optimal multi-group multicast beamforming structure in \eqref{general structure}, we further rewrite it as 
\begin{equation}\label{key}
	\mathbf W^\circ=\left( \mathbf I_L+\mathbf H \bm\Theta^\circ \mathbf H^H  \right)^{-1} \mathbf H\mathbf B^\circ\mathbf P^\circ,
\end{equation}  
where $ \mathbf B^\circ\triangleq \mathrm{blkdiag}\{\mathbf b^\circ_1,\cdots,\mathbf b^\circ_G  \} $ with $ \mathbf b_g^\circ\triangleq [b_{g1}^\circ,\cdots,b_{gK_G}^\circ]^T $ and $ \mathbf P^\circ\triangleq \mathrm{diag}\{\sqrt{p_1^\circ},\cdots,\sqrt{p_G^\circ} \} $. In this form, it is evident that the optimal multi-group multicast beamforming structure consists of the following four parts:
\begin{itemize}
	\item The first part $ \mathbf H $ is a complete channel matrix, which contains channel directions towards all users. These directions are also known as MRT directions.
	\item The second part $ \left( \mathbf I_L+\mathbf H \bm\Theta^\circ \mathbf H^H  \right)^{-1} $ is the inversion of the sum of an identity matrix and a weighted channel covariance matrix. It rotates the MRT directions to reduce the inter-group interference. The parameter $ \theta_{gk}^\circ $ represents the priority assigned to user $ {k} $ in group $ g $, with a lager value indicating that the beamforming vectors of other groups are more orthogonal to the corresponding channel $ \mathbf h_{gk} $. 
	\item The third part  $ \mathbf B^\circ $ is a block-diagonalized coefficient matrix, which is the primary difference between multicast and unicast. Parameter $ b_{gk}^\circ $ represents the priority of user $ k $ in group $ g $, with a larger values indicating that the group beamforming direction $ \mathbf w_g $ is more aligned to $ \mathbf h_{gk} $.  
	\item The fourth part $ \mathbf P^\circ $ is the power allocation matrix  containing the  power allocated to all beamforming vectors $ \{\mathbf w_g \}$.
\end{itemize}

Considering a special case when there is a single user group $g$, i.e., single-group multicast transmission, the corresponding optimal beamforming structure in \eqref{equivalent form} is simplified as $ \mathbf w_g^\circ=\mathbf H_g\mathbf d_g^\circ $. This implies that the second part is an identity matrix and the optimal beamforming solution is determined by directly optimizing the weight vector $\mathbf d_g \in\mathbb{C}^{K_g\times 1}$ instead of the beamforming vector $ \mathbf w_g\in\mathbb{C}^{L\times 1} $. Thus, the optimal weight vector $ \mathbf d_g^\circ $ for maximizing the minimum received signal power $  |\mathbf h_{gk}^H\mathbf w_g|^2 $ is given as
\begin{equation}\label{group channel}
	\mathbf d_g^\circ=\arg\max_{\mathbf d_g} \min_{k\in\mathcal K_g} \left\{ |\mathbf h_{gk}^H\mathbf H_g\mathbf d_g|^2\right\}\,\,\text{s.t.}\,\, \|\mathbf H_g\mathbf d_g\|^2\leq P_t.
\end{equation}
Note that $ \mathbf w^\circ_g=\mathbf H_g\mathbf d_g^\circ $ is the optimal beamforming solution when the number of groups is $ G=1 $, but it is not optimal for multi-group multicast scenarios since the inter-group interference is not considered. 

Regarding multi-group multicast scenarios, the second part  $ (\mathbf I_L+\mathbf H \bm\Theta^\circ \mathbf H^H)^{-1}  $ is required to be considered. This component serves to rotate the group channel matrix $ \mathbf H_g $ into the null space of $\mathbf H_{-g}\triangleq \{ \mathbf h_{11},\cdots,\mathbf h_{g-1,K_{g-1}},\mathbf h_{g+1,K_{g+1}},\cdots,\mathbf h_{GK_G} \}  $, thereby mitigating inter-group interference. In general, it is hard to determine the optimal $ \mathbf \Theta^\circ $ and $ \mathbf D^\circ  $ due to the NP-hard nature of the multicast beamforming design. In Section \ref{Sec:WSR}, we will propose an ultra-low-complexity algorithm based on the optimal multicast beamforming structure to address such issue.

\section{Low-Dimensional Beamforming Structure}
\label{sec:lowDimBF}

\par The primary challenge of finding the optimal beamforming solution to problem \eqref{P4} lies in the undetermined parameter matrices $ \mathbf \Theta^\circ $ and $ \mathbf D^\circ $. Although the optimal parameter matrices $ \mathbf \Theta  $ and $ \mathbf D $ are challenging to calculate, they can be easily attained or even negligible in some asymptotic scenarios, leading to low-complexity and low-dimensional beamforming solutions. In this section, we initially explore two types of low-dimensional structures: one being universal, and the other being asymptotic. We then extend some well-known low-complexity beamforming algorithms to multi-group multicast scenarios, leveraging the asymptotic analysis of the optimal parameter $ \mathbf \Theta^\circ $. Subsequently, we introduce low-dimensional reformulations for the original problems \eqref{P4} and \eqref{P2}.  

\subsection{Range Space (RS) Beamforming}	
In XL-MIMO systems where the number of transmit antennas is much larger than the number of total users, i.e., $ L\gg K $, the computational complexity of calculating the optimal beamforming sharply increases with $L$. Here, we provide a low-dimensional structure to reduce the computational complexity by introducing the following Proposition \ref{low dimension}.   
\begin{proposition}\label{low dimension}
	Any optimal solution $ \mathbf w_g^\circ $ of problem \eqref{P4}   and problem \eqref{P2}  must exist within the range space of the complete channel matrix $ \mathbf H $, i.e., $ \mathbf w_g^\circ=\mathbf H\mathbf a_g^\circ, \forall g\in\mathcal G $, with $\mathbf a_g^\circ\in\mathbb{C}^{K\times 1}  $.
\end{proposition}

\textit{Proof:} By applying matrix identity $ (\mathbf I_L+\mathbf X\mathbf Y)^{-1}\mathbf X=\mathbf X(\mathbf I_{K}+\mathbf Y\mathbf X)^{-1} $ where $ \mathbf X\in\mathbb{C}^{L\times K} $ and $ \mathbf Y\in\mathbb C^{K\times L} $,  any optimal solution shown in \eqref{general structure} can be rewritten as
\begin{equation}\label{low dimensional}
	\mathbf W^\circ =\mathbf H \left(\mathbf I_{K}+\bm\Theta^\circ\mathbf H^H\mathbf H \right)^{-1}\mathbf D^\circ.
\end{equation} 
Denote $ \mathbf A^\circ\triangleq \left(\mathbf I_{K}+\bm\Theta^\circ\mathbf H^H\mathbf H \right)^{-1}\mathbf D^\circ\in\mathbb C^{K\times G} $, we conclude that any optimal beamforming vector for each user group $ g $ must lie in the RS of the complete channel matrix.$ \hfill\blacksquare $ 

Leveraging Proposition \ref{low dimension}, the RS beamforming is given as
 \begin{equation}\label{eq:RS}
 	\mathbf W=\mathbf H\mathbf A,
 \end{equation}
where $ \mathbf A\in\mathbb C^{K\times G} $ has a lower dimension compared to $\mathbf W$. Notably, the dimension of $ \mathbf A$ is independent of the number of transmit antennas $L$. It implies that substituting $\mathbf W$ with $\mathbf H\mathbf A$ in problem \eqref{P4} and \eqref{P2} significantly reduces the optimization dimension of the beamforming matrix. Further elaboration will be provided in Section \ref{sec:LDref}.
\par 
In the following subsections, we discover more low-dimensional  beamforming structures based on the asymptotic analysis of the optimal parameter $ \mathbf \Theta^\circ$ in (\ref{low dimensional}).

\subsection{MRT Beamforming}
 In the low SNR regime, i.e., $\sigma_{gk}^2\rightarrow \infty$ (therefore $ \theta_{gk}=\frac{\lambda_{gk}}{\sigma_{gk}^2}\rightarrow 0 $), the system is noise-limited and the beamforming matrix in \eqref{low dimensional} converges to
\begin{equation}\label{low SNR}
\lim\limits_{P_t\rightarrow 0}	\mathbf W^\circ =\mathbf H\mathbf D^\circ,
\end{equation} 
where the inversion part converges to the identity matrix and $ \mathbf D^\circ $ includes both the asymptotic power allocation and coefficients of the linear combination for the group-channel direction. It implies that MRT beamforming achieves a good performance in the low SNR regime. Thus, a natural extension of the well-known MRT beamforming in multi-group multicast transmission is given as
\begin{equation}\label{MRT}
	\mathbf W=\mathbf H\mathbf D,
\end{equation} where $ \mathbf D\triangleq \mathrm{blkdiag}\{\mathbf{d}_1,\cdots,\mathbf d_G\}\in\mathbb {C}^{K\times G} $ with $ \mathbf d_g\triangleq [d_{g1},\cdots,d_{gK_G}]^T $. It differs from Proposition \ref{low dimension} since $ \mathbf D $ is a block diagonal matrix containing $ K $ variables while $ \mathbf A $ is a full matrix containing $ K\times G $ variables. This strategy maximizes the minimum received signal power $  |\mathbf h_{gk}^H\mathbf w_g|^2 $ received at group $ g $ while ignoring the interference from other user groups. 

\subsection{ZF-based and Regularized ZF-based Beamforming}
When SNR is high, i.e., $ P_t\rightarrow \infty $, the system is in the interference-limited region. We focus on the case $ L\geq K $ with at least one spatial degree-of-freedom per user. In this scenario, each parameter $ \theta_{gk} $ tends to infinity and  \eqref{low dimensional}
converges to
\begin{equation}\label{zf}
\lim\limits_{P_t\rightarrow \infty}	\mathbf W^\circ=\mathbf H(\mathbf H^H\mathbf H)^{-1}\mathbf \Theta^{\circ^{-1}}\mathbf D^\circ.   
\end{equation}
The extension of ZF beamforming in multi-group multicast transmission is
 \begin{equation}\label{MC ZFBF}
 	\mathbf W=\mathbf H\left(\mathbf H^H\mathbf H\right)^{-1} \mathbf D.
 \end{equation}
 
\par 
Similar to the unicast-only transmission, to achieve numerical stability and robustness to channel uncertainty, regularized ZF (RZF) beamforming is usually considered by forcing $ \sum_{g=1}^{G}\sum_{k=1}^{K_g}\theta_{gk}=P_t $ \cite{peel2005vector}. This leads to the following RZF beamforming
\begin{equation}\label{rzf}
	\mathbf W=\mathbf H \left(\frac{1}{P_t}\mathbf I_{K}+\mathbf H^H\mathbf H\right)^{-1}\mathbf D.
\end{equation}

\subsection{Multicast ZF and RZF-based Beamforming}
Recall that the optimal multi-group multicast beamforming has an equivalent form \eqref{equivalent form}, from which we obtain
\begin{equation}\label{multicast zf}
	\lim\limits_{P_t\rightarrow \infty}	\mathbf w_g^\circ=\left(\mathbf H_{-g}\mathbf H_{-g}^H\right)^{\dagger}\mathbf H_g \tilde{\mathbf d}_g^\circ, \forall g\in\mathcal G
\end{equation}
where $ \dagger $ denotes the pseudo-inverse of a matrix. From this asymptotic result, we propose the following two useful low-dimensional beamforming structures
\begin{subequations}\label{MZF and MRZF}
	\begin{align}
	\label{MZF}	\mathbf w_g&=\left(\mathbf H_{-g}\mathbf H_{-g}^H\right)^{\dagger}\mathbf H_g \mathbf d_g, \,\,\forall g\in\mathcal G,\\
	\label{MRZF}	\mathbf w_g&=\left(\frac{1}{P_t}\mathbf I_L+\mathbf H_{-g}\mathbf H_{-g}^H\right)^{\dagger}\mathbf H_g \mathbf d_g, \,\,\forall g\in\mathcal G,
	\end{align}
\end{subequations}
which are referred as multicast ZF (MZF) and multicast RZF (MRZF), respectively. 

\begin{remark}\label{difference MZF and ZF}
	Although \eqref{MZF} has a similar structure with \eqref{MC ZFBF}, they have different mathematical implications. As $ P_t\rightarrow \infty $, the matrix $\left(\mathbf H_{-g}\mathbf H_{-g}^H\right)^{\dagger} $ rotates group channel matrix $ \mathbf H_g $ into the null space of the matrix $\mathbf H_{-g}  $, which implies any group-channel matrix $ \mathbf H_g $ satisfies $ \mathbf H_g^H\left(\mathbf H_{-i}\mathbf H_{-i}^H\right)^{\dagger}\mathbf H_i=\mathbf 0, \forall i\neq g, i\in\mathcal G$. Note that $ \mathbf H^H\left[\left(\mathbf H_{-1}\mathbf H_{-1}^H\right)^{\dagger}\mathbf H_1,\cdots,\left(\mathbf H_{-G}\mathbf H_{-G}^H\right)^{\dagger}\mathbf H_G  \right] $ is a block diagonal matrix and therefore the MZF beamforming \eqref{MZF} is sufficient to eliminate the inter-group interference. 
	This contrasts to the matrix inversion $ \left(\mathbf H^H\mathbf H\right)^{-1} $ in the classical ZF beamforming \eqref{MC ZFBF}, which rotates all  channel vectors to be orthogonal to each other, i.e., $ \mathbf H^H \mathbf H \left(\mathbf H^H\mathbf H\right)^{-1} $ results in a diagonal matrix. 
\end{remark}

\subsection{Large-scale MIMO Systems}
Next, we delve into the asymptotic beamforming structure for XL-MIMO when the number of transmit antenna $ L $ goes to infinite. In \eqref{low dimensional}, the value of $ \mathbf H^H\mathbf H $ grows with $ L $, it is obviously that
\begin{equation}\label{key}
	\lim_{L\rightarrow \infty} \mathbf W^\circ=\mathbf H\left(\mathbf H^H\mathbf H\right)^{-1}\mathbf \Theta^{\circ^{-1}}\mathbf D^\circ, 
\end{equation}
which implies that the ZF-based beamforming is asymptotically optimal when $ L $ goes to infinity. 

A similar low dimensional structure has been proposed in \cite{Dong2020} and adopted by \cite{Dong2020,Zhang2022MMF,Shadi2022,Zhang2023}. This approach employs an asymptotic fixed-point iteration to directly calculate the rotated channel matrix $ \widetilde{\mathbf H}=\mathbf H(\mathbf I_{K}+\mathbf \Theta\mathbf H^H\mathbf H)^{-1} $ and then optimize $ \mathbf D $. It should be noted that the proposed approach in \cite{Dong2020} is only asymptotically optimal when $ L\rightarrow \infty $. But it is tailored for the QoS problem and cannot be extended to solve the general utility problem \eqref{P4}.

\subsection{Low-dimensional Reformulations}
\label{sec:LDref}
The low-dimensional structures introduced in the previous subsections are advantageous for reducing the computational complexity in beamforming design by removing the dependence of the beamforming dimension on the number of transmit antenna $ L $. Here, we take the RS structure (\ref{eq:RS}) as an example to show its benefits. To be specific, by replacing the original high-dimensional beamforming matrix, i.e., $ \mathbf W\in\mathbb{C}^{L\times G} $,  with the low-dimensional RS beamforming matrix i.e., $ \mathbf H\mathbf A $, the original problem \eqref{P4} and \eqref{P2} can be respectively reformulated as
\begin{subequations}
	\label{low dimensional general function}
	\begin{align}
		\max\limits_{\mathbf A}\,\, & f( {R}_1,\cdots,{R}_G)\\
		\text{s.t.}\,\,& \mathrm{Tr}(\mathbf A\mathbf A^H\mathbf F)\leq P_t,
	\end{align}
\end{subequations}
where 
\begin{equation}\label{key}
	{R}_g=\min_{k\in\mathcal K_g} \log\left(1+\frac{|\mathbf f_{gk}^H\mathbf a_g|^2}{\sum_{i=1,i\neq g}^G|\mathbf f_{gk}^H\mathbf a_i|^2+  \sigma_{gk}^2\!}\right),
\end{equation}
and 
\begin{subequations}
	\label{P3}
	\begin{align}
		\min\limits_{\mathbf A}\,\, & \mathrm{Tr}(\mathbf A\mathbf A^H\mathbf F)\\
		\text{s.t.}\,\,& \frac{|\mathbf f_{gk}^H\mathbf a_g|^2}{\sum_{i=1,i\neq g}^G|\mathbf f_{gk}^H\mathbf a_i|^2+  \sigma_{gk}^2\!}\geq \alpha_{g}, \forall k\in\mathcal K_g,\forall g\in\mathcal G,
	\end{align}
\end{subequations}
where $ \mathbf F\triangleq [\mathbf f_{11},\cdots,\mathbf f_{GK_G}]\in\mathbb{C}^{K\times K} $ with $ \mathbf f_{gk}=\mathbf H^H\mathbf h_{gk} $ and $ \mathbf A\triangleq[\mathbf a_1,\cdots,\mathbf a_G]\in\mathbb{C}^{K\times G} $. For both reformulated problems, the dimension of the optimization variables decreases from $ L\times G $ to $ K\times G $ and the complexity of matrix inversion decreases from $ \mathcal O(L^3) $ to $ \mathcal O(K^3) $. 


\section{Efficient Beamforming Algorithms for WSR Maximization}\label{Sec:WSR}
 In this section, we take WSR as an instance of the general utility function, i.e., 
$f\left(R_1,\cdots,R_G\right)=\sum_{g=1}^{G} \zeta_g R_g$,  where $ \zeta_g$ refers to the weight of user group $ g $. Our focus is on deriving efficient beamforming algorithms to solve the following WSR maximization problem:
 \begin{equation}\label{WSR}
 	\max_{\mathbf W}\,\	\sum_{g=1}^{G} \zeta_g R_g, \,\, \text{s.t.} \,\,\mathrm{Tr}(\mathbf W\mathbf W^H)\leq P_t.
 \end{equation}
where  $ R_g=\min_{k\in\mathcal K_g} \left \{\log\left(1+\gamma_{gk}\right) \right \} $. This problem is more difficult to solve than the MMF problem when $f\left(R_1,\cdots,R_G\right)=\min_{g\in \mathcal{G}} R_g$, since the optimal rate targets varies across distinct user groups. Utilizing SIT duality requires a $G$-dimensional exhaustive search over the optimal rate targets, leading to
exponential computational complexity, making it impractical to solve such problem. Generally, problem (\ref{WSR}) has two primary challenges: 
\begin{enumerate}
    \item Non-convex SINR expressions for all users; 
    \item Non-smooth rate expressions for all multicast streams.
\end{enumerate}
The first challenge has been extensively studied in unicast-only
transmission, classical algorithms such as WMMSE, FP, and WSR-MM have emerged to address the non-convexity of SINR expressions. 
The second challenge posed by non-smooth max-min rate expressions has led to the introduction of some approaches, i.e., linearization \cite{asghari2022transformation},  subgradient ascent (SA) algorithm \cite{boyd2003subgradient}, and the LogSumExp (LSE)-based algorithm \cite{xu2001smoothing}. These approaches are briefly introduced below:
\begin{itemize}
    \item Linearization involves introducing auxiliary variables to substitute the max-min rates in the objective function  and adding additional rate constraints for all user groups to reformulate the problem \cite{asghari2022transformation}. Such approach inevitably increases the optimization dimension. 
    \item The SA algorithm \cite{boyd2003subgradient} is a typical approach for solving non-smooth problems. Based on a certain step size, it iteratively updates the solution towards the subgradient direction of the non-smooth objective function until convergence. However, choosing an appropriate rule to update the step size is challenging and can significantly impact the convergence speed.
    \item The LSE method approximates the non-smooth objective function using the  LSE function \cite{xu2001smoothing}. For example, the LSE of $R_g$ is $\mathrm{LSE}_g = -\mu\log\left(\sum_{k=1}^{K_g} \exp({-R_{gk}}/{\mu}) \right)$, where $R_{gk}=\log\left(1+\gamma_{gk}\right)$. However, choosing a proper value for $ \mu $ is challenging. 
    The complicated LSE function introduces extra challenges in solving the original problem, and the approximation error prohibits the identification of the optimal beamforming structure.
\end{itemize}

 \par
Due to the aforementioned limitations of existing algorithms, we next propose a novel  optimization algorithm to solve the non-convex non-smooth WSR problem (\ref{WSR}) based on the optimal and low-complexity beamforming structures we introduced in Section \ref{general optimal} and Section \ref{sec:lowDimBF}.

\subsection{Problem Reformulation}
To better characterize the multi-group multicast beamforming, we move the power constraint into the SINR expression leading to the following unconstrained problem:
\begin{equation}\label{unc WSR}
	\max_{\mathbf W} \sum_{g=1}^G \zeta_g\min_{k\in \mathcal{K}_g}\left\{\log\left(1+\widehat{\gamma}_{gk} \right )\right\},
\end{equation}
where
\begin{equation}\label{key}
	\widehat{\gamma}_{gk}\triangleq\frac{|\mathbf h_{gk}^H\mathbf w_g|^2}{\sum_{i\neq g} |\mathbf h_{gk}^H \mathbf w_i|^2+\frac{\sigma_{gk}^2}{P_t} \mathrm{Tr}\left(\mathbf W\mathbf W^H\right) }.
\end{equation} 
The relation between \eqref{WSR} and $\eqref{unc WSR} $ is provided in the following Proposition \ref{remove power}.
\begin{proposition}
	\label{remove power}
	For any locally optimal solution $ \mathbf W^\diamond $ of problem \eqref{WSR}, there exists a corresponding locally optimal solution $ \mathbf W^\ddagger $ of the unconstrained problem $ \eqref{unc WSR} $ such that $ \mathbf W^\diamond=\sqrt{\frac{P_t}{\mathrm{Tr}(\mathbf W^\ddagger\mathbf W^{\ddagger^H})}} \mathbf W^\ddagger  $. 
\end{proposition}
\textit{Proof:} The detail proof follows the procedure in \cite{Zhao2023}. $ \hfill\blacksquare $ 

  Leveraging Proposition \ref{remove power}, we can efficiently solve problem \eqref{WSR} by directing our focus towards solving the unconstrained problem \eqref{unc WSR} in the following.
\subsection{Cyclic Maximization Method}\label{FPM}
\par In this subsection, we introduce a cyclic maximization (CM)-based method to solve problem \eqref{unc WSR}. The main idea of CM is to construct a high dimensional surrogate objective function and then solve each block cyclically to obtain a stationary point of the original non-convex problem \cite{journee2010generalized,Sun2017Major}.
We start with introducing the following Proposition \ref{lemma CM1} to transform \eqref{unc WSR} into a more tractable form.

\begin{proposition}\label{lemma CM1}
	By introducing auxiliary variables $ \{\xi_{gk}, \eta_{gk} \}$, problem \eqref{unc WSR} can be equivalently reformulated as
\begin{equation}\label{unc reWSR}
	\max_{\mathbf W, \bm\xi,\bm\eta}\,\, \sum_{g=1}^G \zeta_g \min_{k\in \mathcal{K}_g} \{ h_{gk}(\mathbf W,\xi_{gk},\eta_{gk}) \}, 
\end{equation}
where $ \bm\xi\triangleq[\xi_{11},\cdots,\xi_{GK_G}]^T, \bm\eta\triangleq[\eta_{11},\cdots,\eta_{GK_G}]^T $, and  
\begin{equation}	\label{rate lower bound}
	\begin{split}
		h_{gk}(\mathbf W,&\xi_{gk},\eta_{gk})\!\triangleq\!\log(1\!+\!\xi_{gk})\!+\!2\sqrt{1\!+\!\xi_{gk}}\Re\{\eta_{gk}^H\mathbf h_{gk}^H\mathbf w_g  \}\!\\
	&-\!|\eta_{gk}|^2\left(\sum_{i=1}^G\!|\mathbf h_{gk}^H\mathbf w_i|^2+\frac{\sigma_{gk}^2}{P_t} \mathrm{Tr}(\mathbf W\mathbf W^H)\!\!\right)\!\!-\!\xi_{gk}.
\end{split}
\end{equation}
\end{proposition} 
\textit{Proof:} The convex function $ \log(\frac{1}{x}) $ with $ x\in\mathbb R_+ $ has the following lower bound
\begin{equation}\label{log lower bound}
	 \log\left(\frac{1}{x}\right)\geq \log\left(\frac{1}{x_0}\right)-\frac{1}{x_0}(x-x_0)  
\end{equation}
with equality achieved at $ x_0=x $. By plugging $ x=\frac{1}{1+\widehat{\gamma}_{gk}} $ and $ x_0=\frac{1}{1+\xi_{gk}} $ into \eqref{log lower bound}, we obtain the following surrogate function \begin{equation}\label{key}
\begin{split}
		&\log(1+\hat{\gamma}_{gk})\geq \log(1+\xi_{gk})-\frac{1+\xi_{gk}}{1+\widehat{\gamma}_{gk}}+1\\
		&\geq \log(1+\xi_{gk})-\xi_{gk}+\frac{(1+\xi_{gk})|\mathbf h_{gk}^H\mathbf w_g|^2}{\sum_{i=1}^G |\mathbf h_{gk}^H\mathbf w_i|^2+\frac{\sigma_{gk}^2}{P_t} \mathrm{Tr}(\mathbf W\mathbf W^H)}\notag
\end{split}
\end{equation} with equality holding when $ \xi_{gk}=\widehat{\gamma}_{gk}. $ Further, the convex fractional function $ \frac{|x|^2}{y} $ with $ x\in\mathbb{C} $ and $ y\in\mathbb R_+ $ can be lower bounded by its first order Taylor expansion as 
\begin{equation}\label{QT}
	\frac{|x|^2}{y}\geq 2\Re\left\{\frac{x_0^H}{y_0}x\right \}-\frac{|x_0|^2}{y_0^2}y
\end{equation}
with equality achieved at $ (x_0,y_0)=(x,y) $. By substituting $ x=\sqrt{1+\xi_{gk}}\mathbf h_{gk}^H\mathbf w_g, y=\sum_{i=1}^G |\mathbf h_{gk}^H\mathbf w_i|^2+\frac{\sigma_{gk}^2}{P_t} \mathrm{Tr}(\mathbf W\mathbf W^H) $ and $ \eta_{gk}=x_0/y_0$ into \eqref{QT}, we obtain the two-layer surrogate function $ h_{gk}(\mathbf W,\xi_{gk},\eta_{gk}) $ in \eqref{rate lower bound}. $ \hfill\blacksquare $

\par Problem \eqref{unc reWSR} remains non-convex, but it is obvious that, it can be solved by the CM method with the following updates
\vspace{-2cm}
\begin{subequations}
\begin{align}
\label{updat xi}	&\xi_{gk}^{[t+1]}=\gamma_{gk}^{[t]},\\
\label{updat eta} &\eta_{gk}^{[t+1]}=\frac{\sqrt{1+\xi_{gk}^{[t+1]}} \mathbf h_{gk}^H\mathbf w_g^{[t]}  }{\sum_{i=1}^G \left|\mathbf h_{gk}^H\mathbf w_i^{[t]}\right|^2+\frac{\sigma_{gk}^2}{P_t} \mathrm{Tr}\left(\mathbf W^{[t]}\mathbf W^{[t]^H}\right)},\\
\label{subproblem over W}&\mathbf W^{[t+1]}=\!\arg\max \! \sum_{g=1}^G\! \zeta_g \!\min_{k\in \mathcal{K}_g} \left\{ h_{gk}\left(\mathbf W \!,\xi_{gk}^{[t+1]}\!,\eta_{gk}^{[t+1]}\right) \!\right\}.
\end{align}
\end{subequations}

\par For given $ \{ \bm\xi,\bm\eta\} $, the subproblem \eqref{subproblem over W} with respect to $ \mathbf W$ is a non-smooth convex quadratic programming and thus can be directly solved using standard convex optimization approaches (i.e., interior-point methods) implemented by a certain solver (i.e., SeDuMi) in the CVX toolbox \cite{grant2014cvx}. 
Such approach of using CM to solve problem (\ref{unc reWSR}) as well as using the standard solvers in CVX to solve problem \eqref{SOCP} is referred as the standard CM method as summarized in Algorithm \ref{alg:AO}. 
\par The equivalence between the CM framework and the classical MM theory has been confirmed in \cite{Sun2017Major}. Besides, the recent study \cite{zhang2023discerning} has established the equivalence between WMMSE, FP, and MM for solving the WSR problems. Therefore, we could infer that algorithms using convex approximation and employing a standard optimization solver in CVX \cite{grant2014cvx} to solve each convex subproblem  exhibit comparable performance.
\setlength{\textfloatsep}{7pt}	
\begin{algorithm}[t!]
	\caption{Standard CM framework for problem \eqref{WSR}}
	\label{alg:AO}
	\textbf{Initilize:} Set initial feasible $ \mathbf W^{[0]} $; set $ t=1 $.\;
	\Repeat{convergence}{ 
		$ t\leftarrow t+1 $\;
		Update the auxiliary variables $ \xi_{gk}^{[t]} $ and $ \eta_{gk}^{[t]} $ by \eqref{updat xi} and \eqref{updat eta} for $  k\in\mathcal{K}_g,\forall g\in\mathcal G $\;
		Update $ \mathbf W^{[t]}$ by solving problem (\ref{SOCP}) optimally\;}
	Scale and output $ \mathbf W^{\diamond}=\sqrt{\frac{P_t}{\mathrm{Tr}(\mathbf W^{[t]}\mathbf W^{[t]^H})}} \mathbf W^{[t]}$.
\end{algorithm}

\vspace{-1mm}
\subsection{Proposed Efficient Optimization Algorithms}

\par Although the standard CM algorithm in Algorithm \ref{alg:AO} successfully solves subproblem \eqref{subproblem over W} using the standard CVX toolbox, this approach is not cost-effective due to the substantial time occupied to parse and canonicalize the original subproblem into a standard form for CVX solvers to understand. Moreover, problem \eqref{subproblem over W} is non-differentiable at the points where two or more functions of $ \{h_{gk}\}$ share the same value. To avoid these limitations of solving \eqref{subproblem over W} directly using CVX, we propose a novel optimization algorithm based on its optimal beamforming structure. Specifically, by introducing a set of slack auxiliary variables $ \mathbf z=[z_1,\cdots,z_G]^T $, we aim to solve the following equivalent problem of \eqref{subproblem over W}:
\begin{subequations}
\vspace{-1mm}
	\label{SOCP}
	\begin{align}
		\max\limits_{\mathbf W,\mathbf z}\,\, &\sum_{g=1}^G\zeta_gz_g \\
		\label{SOCP:C1}		\text{s.t.}\,\,& z_g\leq h_{gk}(\mathbf W \!,\xi_{gk},\eta_{gk}),\forall g\in\mathcal G,\forall k\in\mathcal K_g.
	\end{align}
\end{subequations}
Problem \eqref{SOCP} is a smooth convex SOCP problem and can be efficiently solved via its Lagrange dual problem. The Lagrangian function of \eqref{SOCP} is given as
\vspace{-2mm}
\begin{equation*}
	\mathcal L_{\eqref{SOCP}}(\bm\delta,\mathbf W,\mathbf z)\triangleq\sum_{g=1}^G\zeta_gz_g- \sum_{g=1}^G\sum_{k=1}^{K_g} \delta_{gk}\!\left(z_g\!-\!h_{gk}\!\left(\mathbf W,\xi_{gk},\eta_{gk}\right)\right),
\end{equation*} 
where $ \delta_{gk}\geq 0 $ is the dual variable corresponding to constraint \eqref{SOCP:C1}, and $ \bm\delta\triangleq [\bm\delta_1,\cdots,\bm\delta_G] $ with $ \bm\delta_g\triangleq [\delta_{g1},\cdots,\delta_{gK_g}]^T $. The first-order derivatives of the Lagrange function $ \mathcal L_{\eqref{SOCP}}(\bm\delta,\mathbf W,\mathbf z) $ with respect to $ z_g$ and $ \mathbf w_g $ are respectively given as
\vspace{-1mm}
\begin{equation*}\label{first order}
	\begin{aligned}
		\frac{\partial \mathcal L_{\eqref{SOCP}}}{\partial z_g}&\!=\zeta_g-\sum_{k=1}^{K_g} \delta_{gk},\\
		\frac{\partial \mathcal L_{\eqref{SOCP}}}{\partial \mathbf w_g}&\!=\!\sum_{k=1}^{K_g}\!2\acute{d}_{gk} \mathbf h_{gk} \!-\!\sum_{i=1}^G\! \sum_{k=1}^{K_i}\!2\left(\acute{\theta}_{ik}\mathbf h_{ik}  \mathbf h_{ik}^H\!+\!\frac{\delta_{ik}\sigma_{ik}^2}{P_t}\mathbf I_L\right)\mathbf w_g,\\ 
	\end{aligned}
\end{equation*}
where $ \acute{d}_{gk},\forall g\in\mathcal G,k\in\mathcal K_g$, and $  \acute{\theta}_{ik}, \forall i\in\mathcal G,k\in\mathcal K_i$
are respectively defined as
\begin{equation}\label{key}
	\begin{aligned}
		\acute{d}_{gk}&\triangleq \delta_{gk}\eta_{gk}\sqrt{1+\xi_{gk}},&
		\acute{\theta}_{ik}&\triangleq  \delta_{ik} |\eta_{ik}|^2.\\
	\end{aligned}
\end{equation}
\par Since \eqref{SOCP} is convex and strictly feasible, it satisfies the Slater's condition and the strong duality holds \cite{boyd2004convex}. Therefore, the optimal solution of \eqref{SOCP}, together with the optimal Lagrange dual variable, satisfies the following KKT conditions
\vspace{-3mm}
\begin{subequations}
	\begin{align}
\label{kkt for z_g}&\zeta_g-\sum_{k=1}^{K_g} \delta_{gk}=0,g\in\mathcal{G},\\
\label{kkt for w_g}&\sum_{k=1}^{K_g}\!2\acute{d}_{gk} \mathbf h_{gk}\! -\!\sum_{i=1}^G\! \sum_{k=1}^{K_i}\!2\!\left(\!\acute{\theta}_{ik}\mathbf h_{ik}  \mathbf h_{ik}^H\!+\!\frac{\delta_{ik}\sigma_{ik}^2}{P_t}\mathbf I_L\!\right)\!\mathbf w_g=\mathbf 0,g\in\mathcal{G},\\ 
\label{complemetntary}&\delta_{gk}(z_g-h_{gk})=0,g\in\mathcal{G},k\in\mathcal K_g,
	\end{align}
\end{subequations}
where \eqref{kkt for z_g} and \eqref{kkt for w_g} are the first-order stationary conditions, and \eqref{complemetntary} refers to the complementary slackness conditions. The primal and dual feasibility conditions are omitted here.

\par According to \eqref{SOCP}, we could directly obtain the optimal slackness variable $ z_g $ as $ z_g^\star=\min_{k\in \mathcal{K}_g}\{h_{gk} \}. \forall g\in\mathcal G $. Moreover, from \eqref{kkt for w_g}, we further reveal the optimal beamforming solution structure for problem \eqref{SOCP} in Theorem \ref{Pro: Optimal beamforming of subproblem}.
\begin{theorem}\label{Pro: Optimal beamforming of subproblem}
	The optimal beamforming solution for problem \eqref{SOCP} is given by
\begin{equation}
		\label{up W}	\mathbf w_g^{\star}\!=\left(\mathbf H\acute{\bm\Theta}^\star \mathbf H^H + S_{\sigma}\mathbf I\right)^{-1}\mathbf H_g\acute{\mathbf d}_g^\star, \forall g\in\mathcal G,
\end{equation}
where $ \{\delta_{gk}^\star\} $ are the optimal dual variables for problem \eqref{SOCP}, $ \acute{\bm\Theta}^\star\triangleq \mathrm{blkdiag}\{\acute{\bm\Theta}_1^\star,\cdots, \acute{\bm\Theta}_G^\star \}  $ with $ \acute{\bm\Theta}_g^\star=\mathrm{diag}\{ \acute{\theta}_{g1}^\star,\cdots,\acute{\theta}_{gK_G}^\star  \} $ with $ \acute{\theta}_{gk}^\star=\delta_{gk}^\star|\eta_{gk}|^2 $, $ S_{\sigma}\triangleq \sum_{i=1}^G \sum_{k=1}^{K_i}\frac{\delta_{ik}^\star\sigma_{ik}^2}{P_t} $, and $  \acute{\mathbf d}_g^\star=[\acute{d}_{g1}^\star,\cdots,\acute{ d}_{gK_g}^\star]^T $ with $ \acute{d}_{gk}^\star= \delta_{gk}^\star\eta_{gk}\sqrt{1+\xi_{gk}} $. 
\end{theorem}
\par
It is clear that the beamforming structure \eqref{up W} shares the same structure with the optimal multi-group multicast beamforming structure \eqref{general structure}.
Substituting \eqref{up W} and \eqref{kkt for z_g} into $ \mathcal L_{\eqref{SOCP}}(\bm\delta,\mathbf W,\mathbf z) $, the Lagrange dual problem of \eqref{SOCP} is given by
\begin{equation}\label{dual problem}
	\min_{\bm\delta\in\mathcal H} \,\, \sum_{g=1}^G\sum_{k=1}^{K_g} \delta_{gk}h_{gk}(\mathbf W^\star,\xi_{gk},\eta_{gk}),
\end{equation}
where $ \mathcal H\triangleq \mathcal H_1\times \cdots\times\mathcal H_G $ with $ \mathcal H_g\triangleq \{ \delta_{gk}\geq 0: \sum_{k=1}^{K_g} \delta_{gk}=\zeta_g \} $ and $ \mathbf W^\star\triangleq [\mathbf w^\star_1,\cdots,\mathbf w_G^\star] $. The dual problem is convex and the feasible space $ \bm\delta\in\mathcal H $ implies that each dual vector $ \bm\delta_g $ lies in the hyperplane $ \mathcal H_g $. This motivates us to solve problem \eqref{dual problem} base on the projected adaptive gradient descent (PAGD) algorithm. 

\par PAGD is an optimization algorithm that minimize a function iteratively by updating the solution towards the opposite direction of the gradient with an adaptive step size in each iterative. It then includes a projection step to project the updated solution onto the feasible set.  In this work, we propose to solve problem \eqref{dual problem} based on the following updating procedure in each iteration $ [j] $:
\vspace{-2mm}
\begin{subequations}\label{up dual delta}
\begin{align}
\!\!&\bar{\delta}_{gk}^{[j+1]}=\delta_{gk}^{[j]}-\tau_{gk}^{[j]}\!\left(\!h_{gk}^{[j]}\!- \!\min_{i\in\mathcal K_g} \left\{h_{gi}^{[j]}\right\}\! \right)\!,g\in\mathcal G,k\in\mathcal K_g, \\
\label{hyperplane} \!\!&\bm\delta_g^{[j+1]}=\bm \Pi_{\mathcal H_g}(\bar{\bm\delta}_g^{[j+1]}), \forall g\in\mathcal G,
\end{align}
\end{subequations}
where  $ \bar{\bm\delta}_g\triangleq [ \bar{\delta}_{g1},\cdots,\bar{\delta}_{gK_g} ]^T $ refers to the intermediate updates before projection, and $ \tau_{gk}^{[j]} $ is the step size, given by
\vspace{-2mm}
\begin{equation}\label{step size}
	\tau_{gk}^{[j]}=\frac{\delta_{gk}^{[j]}}{h_{gk}^{[j]}- \!\min_{i\in\mathcal K_g}\left\{h_{gi}^{[j]}\right\}+\rho_t^{[j]}},
\end{equation}
with an increasing constant number $ \rho_t^{[j]}=\rho_c+\rho_v\cdot j $.
$ \bm \Pi_{\mathcal H_g}(\bar{\bm\delta}_g^{[j+1]}) $ denotes the projection of  $ \bar{\bm\delta}_g^{[j+1]} $ onto the hyperplane $ \mathcal H_g $, which is defined as
\vspace{-2mm}
\begin{equation}\label{projection}
	\bm \Pi_{\mathcal H_g}\left(\bar{\bm\delta}_g^{[j+1]}\right)=\bar{\bm\delta}_g^{[j+1]}-\frac{\sum_{k=1}^{K_gk}\bar{\delta}_g^{[j+1]}-\zeta_g}{K_g}.
\end{equation}
The optimal dual vector $ \bm\delta^\star $ for each convex problem \eqref{SOCP} can be obtained using the proposed PAGD algorithm summarized in Algorithm \ref{alg:PAGD}. Therefore, the subproblem \eqref{subproblem over W} is optimally solved by substituting $ \bm\delta^\star $ into the optimal beamforming solution \eqref{up W}. By employing the CM framework and PAGD to address \eqref{WSR}, we establish a highly efficient algorithm referred to as CM-PAGD. In contrast to the standard CM approach, our proposed algorithm exhibits lower computational complexity and ensures no loss in performance. This will be further demonstrated in the simulation section. 
\setlength{\textfloatsep}{7pt}	
\begin{algorithm}[t!]
	\caption{PAGD algorithm for solving problem \eqref{SOCP}}
	\label{alg:PAGD}
	\textbf{Initilize:} Set initial feasible $  \bm\delta^{[0]}\in\mathcal H $, set $ j=1 $\;
	\Repeat{convergence}{ 
		$ j\leftarrow j+1 $\;
		Update the primal variable $ \mathbf w_g^{[j]} $ by \eqref{up W} for $ \forall g\in\mathcal G $\;
		Update the dual variable $ \bm\delta_g^{[j]} $ by \eqref{up dual delta} for $ \forall g\in\mathcal G $\;}
\end{algorithm}

\begin{remark}
 The step size follows the rule of square summable but not summable \cite{boyd2003subgradient}, which typically follows
 \begin{equation}\label{eq:PAGDupdate}
 	\tau_{gk}^{[j]}=\frac{x}{y+\rho_v\cdot j},
 \end{equation}
where $ x>0 $, $ y\geq 0 $ are problem-specific parameters and $\rho_v  $ is a decreasing factor. To ensure $ \bar{\delta}_{gk}^{[j+1]}\geq 0 $, we have 
\begin{equation}
\tau_{gk}^{[j]}\leq 	\frac{\delta_{gk}^{[j]}}{\!h_{gk}^{[j]}\!- \!\min_{i\in\mathcal K_g} \left\{h_{gi}^{[j]}\right\}\! }.
\end{equation}
Let $x$ and $y$ in \eqref{eq:PAGDupdate} be defined as \ $x= \delta_{gk}^{[j]}$ and $ y=h_{gk}^{[j]}\!- \!\min_{i\in\mathcal K_g}\{h_{gi}^{[j]}\}+\rho_c $, we end up with the proposed step size \eqref{step size}. This step size enables the dual vector $ \bar{\bm{\delta}}_g^{[j+1]} $ within the subspace $ \{\delta_{gk}\geq 0:\sum_{k=1}^{K_g}\delta_{gk}\leq \zeta_g \} $, and it is therefore easy to project it back onto the hyperplane $ \mathcal H_g $ based on the defined projection rule \eqref{projection}.
\end{remark}

\subsection{Low-dimensional Reformulations for Large Scale Systems}
\par 
As mentioned in Section \ref{sec:lowDimBF}, the proposed low-dimensional beamforming structures can be used to reduce the computational complexity of the beamforming design.
Here, we take the RS structure $\mathbf W=\mathbf H\mathbf A $ as an example to show its benefits in solving the WSR problem for XL-MIMO systems. 
By substituting $\mathbf W=\mathbf H\mathbf A $ into \eqref{WSR}, the WSR problem is reformulated as
\begin{subequations}\label{range space WSR}
	\begin{align}
		\max_{\mathbf A}\,\, &\sum_{g=1}^G \zeta_g\min_{k\in \mathcal{K}_g} \!\!\left\{\log\left(1+ \frac{|\mathbf f_{gk}^H\mathbf a_g|^2}{\sum_{i=1,i\neq g}^G|\mathbf f_{gk}^H\mathbf a_i|^2+  \sigma_{gk}^2} \right)  \!\! \right\}   \\
		\text{s.t.}\,\, &\mathrm{Tr}(\mathbf A\mathbf A^H\mathbf F)\leq P_t.
	\end{align}
\end{subequations}
This formulation is equivalent to \eqref{WSR}, but with significantly reduced dimensionality in the optimization variables.  Therefore, we can solve it using the proposed CM-PAGD methods. 

Other low-dimensional reformulations based on MRT, ZF, RZF, MZF, or MRZF follow a similar process as \eqref{range space WSR}. To avoid redundancy, we omit the details of these reformulation here. A comprehensive comparison among different approaches will provided in the following simulation section.

\subsection{Convergence and Computational Complexity Analysis} 

\subsubsection{Convergence Analysis}
The proposed CM-PAGD algorithm consists of two iteration layers. The outer-layer CM framework outlined in Algorithm \ref{alg:AO} is guaranteed to generate a monotonically increasing sequence of objective values for \eqref{WSR}, as proven in \cite{Sun2017Major,zhou2020intelligent,li2022secure}. Regarding the convergence of the inner-layer iteration for computing the optimal dual variables in Algorithm \ref{alg:PAGD}, it is established in \cite{boyd2003subgradient} that the algorithm is guaranteed to converge if the objective function satisfies the Lipschitz condition. This condition is clearly met in our algorithm since the objective function is continuous differentiable over the convex set $ \mathcal H $. For further details,  readers can refer to \cite{boyd2003subgradient}. 
\subsubsection{Computational Complexity Analysis}
The computational complexity of the proposed CM-PAGD algorithm for each iteration is dominated by updating the beamforming matrix (i.e., line 5 of Algorithm \ref{alg:AO}) based on Algorithm \ref{alg:PAGD}. The complexity of Algorithm \ref{alg:PAGD} is dominated by the matrix inversion in line 4, with an order of $ \mathcal O(GL^3) $. The overall complexity order is $\mathcal O(GL^3\log(\epsilon_1^{-1})\log(\epsilon_2^{-1}))  $, where $ \epsilon_1 $ refers to the convergence tolerance of the outer-layer CM framework, and $ \epsilon_2 $ refers to the convergence tolerance of the inner-layer PAGD algorithm.

 
\section{Simulation Results}\label{simulation}\label{Sec:simulation}
 In this section, we evaluate the computational complexity, convergence, and the WSR performance of the proposed CM-PAGD algorithm based on the optimal beamforming structure or other low-dimensional beamforming structures.
\subsection{Simulation Setup}
\par We consider a symmetric multi-group multicast communication network, where $ K_g=K_G, \forall g\in\mathcal G  $. Unless specified otherwise, the default user set consists of $ G=3 $ groups, with $ K_G=4 $ users per group. The channel of user $ k $ is generated i.i.d. as $ \mathbf h_{gk}\sim\mathcal{CN}(\mathbf 0,\mathbf I_L) $ and the noise variance at user $ k $ is set to $ \sigma_{gk}^2=1 $ so that the transmit SNR defined as SNR $\triangleq P_t/\sigma_{gk}^2  $ is numerically equal to the transmit power. For the proposed algorithms, we set the stopping tolerance for both the outer-layer CM framework and inner-layer iterative algorithms as $ \epsilon_1= \epsilon_2=10^{-4}  $. Additional, the constant $ \rho_t^{[j]} $ in \eqref{step size} is set to $ \rho_t^{[j]}=1+0.02\times j  $
for controlling the convergence accuracy of the inner-layer PAGD algorithm. Without loss of generality, we set the weights $ \zeta_1=\zeta_2=\cdots=\zeta_G=1 $ in our simulations. The initialization of the beamforming vectors for the CM framework is based on MRT directions, e.g., $ \mathbf w_g^{[0]}=\sum_{k=1}^{K_g} \mathbf h_{gk}, \forall g\in\mathcal G  $. All simulation results are averaged over $ 100 $ random channel realizations.

\subsection{Baseline Algorithms}



All schemes considered in the simulation are summarized as follows:
\begin{itemize}
	\item \textbf{standard CM}: This is the standard CM framework we introduced in Algorithm \ref{alg:AO}. Each subproblem \eqref{subproblem over W} is solved using CVX toolbox, leading to an overall computational complexity of $ \mathcal O([GL]^{3.5}\log(\epsilon_1^{-1})) $.
	\item \textbf{CM-SA}: This refers to the algorithm of employing the CM framework to solve problem \eqref{unc reWSR}, while using the SA algorithm \cite{boyd2003subgradient} to solve each non-smooth surrogate problem directly. The gradient of worst-case user per group is selected as the subgradient \cite{Zhang2022MMF}. The overall computational complexity is $ \mathcal O(GL^2\log(\epsilon_1^{-1})\log(\epsilon_2^{-2})) $.
	\item \textbf{CM-LSE}: This refers to the algorithm of employing the CM framework to solve problem \eqref{unc reWSR}, while using the LSE algorithm \cite{xu2001smoothing} to approximate the non-smooth objective function for each surrogate problem \cite{zhou2020intelligent}. The approximated convex and smooth problem is then solved by the gradient ascent approach. The corresponding computational complexity is $ \mathcal O(GL^2\log(\epsilon_1^{-1})\log(\epsilon_2^{-1})) $. 
	\item \textbf{CM-PAGD}: This is the algorithm we proposed based on Algorithm \ref{alg:AO} and Algorithm \ref{alg:PAGD}. Specifically, for each subproblem \eqref{subproblem over W} in Algorithm \ref{alg:AO}, instead of using CVX to solve it directly, we propose to use the PAGD algorithm in Algorithm \ref{alg:PAGD} to address it. CM-PAGD is therefore an optimization toolbox-free algorithm, which reduces the computational complexity. The corresponding computational complexity is $ \mathcal O(GL^3\log(\epsilon_1^{-1})\log(\epsilon_2^{-1}))  $.
	\item \textbf{RS CM-PAGD}: This is the algorithm we proposed based on the RS property discovered in \eqref{low dimensional} and CM-PAGD. Specifically, problem \eqref{WSR} is transformed to \eqref{range space WSR} using the RS property. After that, CM-PAGD is employed to solve \eqref{range space WSR}.  The computational complexity of RS CM-PAGD is $ \mathcal O(K^2L+GK^3\log(\epsilon_1^{-1})\log(\epsilon_2^{-1})) $.
	\item \textbf{X CM-PAGD}: This is the algorithm we proposed based on the low dimensional beamforming structure discovered in  \eqref{MRT}/\eqref{zf}/\eqref{rzf}/\eqref{MZF and MRZF} and CM-PAGD. Specifically, problem \eqref{WSR} is first transformed based on \eqref{MRT}/\eqref{zf}/\eqref{rzf}/\eqref{MZF}/\eqref{MRZF}, respectively for the scenarios of X=MRT/ZF/RZF/MZF/MRZF. Subsequently, CM-PAGD is employed to solve the corresponding transformed problem. The computational complexity of X CM-PAGD is $ \mathcal O(K^2L+GK_g^3\log(\epsilon_1^{-1})\log(\epsilon_2^{-1})) $.
\end{itemize}
 All algorithms, excluding  X CM-PAGD, aim at calculating sub-optimal beamforming solutions for the WSR problem \eqref{WSR}. In contrast, X CM-PAGD algorithms are asymptotically optimal in different regimes. They notably reduce the computational complexity, but may lead to performance degradation in certain regimes. 

\begin{figure}[t!]
	\centering
	\subfigure[Convergence of the outer layer]{\label{Fig conver out}
	\includegraphics[width=0.5\linewidth]{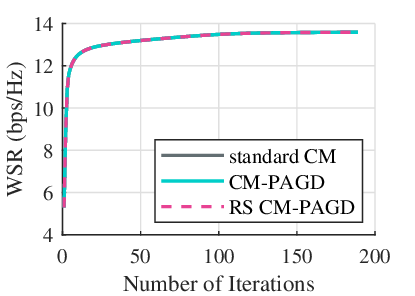}
}\subfigure[Convergence of the inner layer]{\label{Fig conver inner}
		\includegraphics[width=0.5\linewidth]{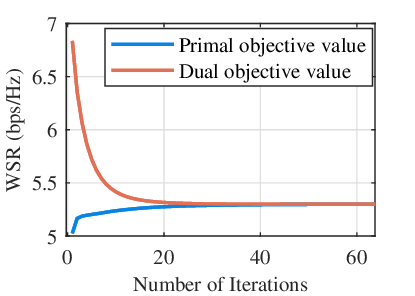}
}
	\caption{Convergence of the proposed algorithms for a certain channel realization.}
	\label{Fig 1}
\end{figure}
\subsection{Convergence of the Proposed Algorithms}
\par  We first check the convergence behavior of the standard CM algorithm, the proposed CM-PAGD and  RS CM-PAGD algorithms. Both CM-PAGD and RS CM-PAGD have two iteration layers, namely, one outer layer for CM and one inner layer for PAGD. The convergence to both iteration layers is illustrated in Fig.$\, \ref{Fig conver out} $ and Fig.$ \,\ref{Fig conver inner} $, respectively. Fig.$\, \ref{Fig conver out} $ shows the WSR performance of all three schemes as the number of  iterations in the outer layer increases when $ L=16 $, SNR=$ 20 $dB.  We would observe that the convergence path of the proposed CM-PAGD and RS CM-PAGD nearly overlap with the standard CM algorithm. This is because the convex surrogate problem \eqref{subproblem over W} is optimally solved by the derived optimal beamforming solution structure \eqref{up W} together with the PAGD algorithm that successfully calculates the Lagrange dual variables. Fig. $ \,\ref{Fig conver inner} $ illustrates the WSR versus the number of iterations in the inner layer for both dual and primal problems. It is obviously that the duality gap between the primal and the dual objective values converges to zero when solving each subproblem \eqref{subproblem over W}.   
\begin{table*}
	\begin{center}
		\caption{Weighted Sum Rate (bps/Hz) Versus Transmit SNR (dB) Comparison among Different Strategies.}
		\label{tab1}
		\begin{tabular}{cccccccccc}
			\toprule
			 SNR &-10 &   -5& 0 & 5 & 10&15&20&25&30 \\
			\hline\hline
			standard CM& 0.5188&1.2833&2.6875 & 4.7814&7.4543 & 10.5215&13.7879&17.1312&20.3602\\
			\hline
			CM-PAGD & 0.5190&1.2832&2.6854 & 4.7792&7.4536 & 10.5143&13.7855&17.1169&20.3101\\
			\hline
			CM-SA& 0.5173&1.2787&2.6758 & 4.7473&7.3766 & 10.3571&13.3432&16.4204&19.5964\\
			\hline 
			CM-LSE& 0.5054&1.2645&2.6717 & 4.7515&7.4074 & 10.4447&13.4231&16.6817&19.8555\\
			\hline
			RS CM-PAGD& 0.5190&1.2832&2.6854 & 4.7792&7.4536 & 10.5143&13.7854&17.1266&20.4048\\
			\bottomrule
		\end{tabular}
	\end{center}
\end{table*}
\subsection{Comparison among the Sub-optimal Algorithms}
\par
\begin{figure}[t]
			\centering
		\includegraphics[width=0.8\linewidth]{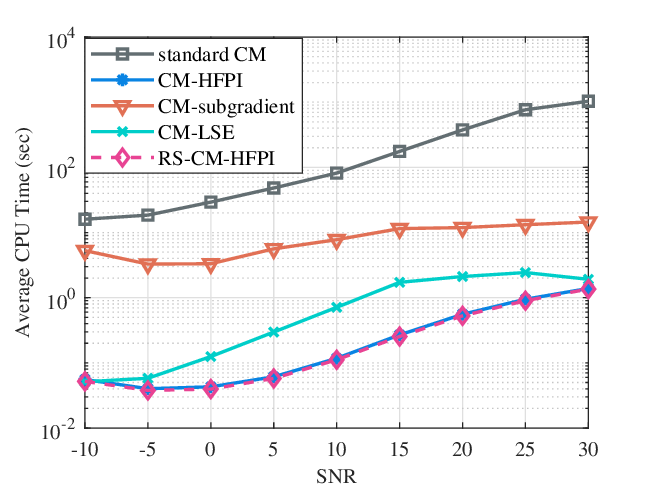}

	\caption{Averaged CPU time versus the transmit SNR for different algorithms when $ L=16 $, $G=3$ and $K_G=4 $.}
	\label{img:performance_vs_SNR}
\end{figure}

 Table.\,\ref{tab1} shows the WSR performance of the five sub-optimal algorithms as the transmit SNR increases from -10dB to 30dB. The number of transmit antenna is fixed to $ L=16 $. Notably, the proposed CM-PAGD and RS-PAGD algorithms solve the WSR problem effectively without any performance loss compared to the standard CM. In comparison, CM-SA and CM-LSE cause certain performance degradation especially when SNR is large. This is due to the fact that in the high SNR regime, the CM-SA algorithm is more prone to oscillate in the vicinity of the non-differentiable optimal point during the inner iteration. Additionally, the LSE approximation function introduces a lager approximation error for the CM-LSE algorithm under these conditions.  Fig.$\, \ref{img:performance_vs_SNR} $ illustrates the corresponding average CPU time versus the transmit SNR. Compared with other baselines schemes, the proposed algorithms exhibit substantial reduction in average CPU time. Notably, CM-PAGD and RS CM-PAGD achieve at lease a 99.65\% time decrease over standard CM across all SNR regimes, highlighting their effectiveness in achieving excellent WSR performance with lower computational complexity. The CPU time consumption for CM-PAGD and RS CM-PAGD is comparable since the number of transmit antennas $ L=16 $ is of the same order as the total number of users $ K=12 $, resulting in a marginal CPU time reduction for RS CM-PAGD. Further more, despite the potential higher complexity per iteration for CM-PAGD and RS CM-PAGD compared to CM-SA and CM-LSE, they exhibit faster convergence rates since CM-SA and CM-LSE converge in the primal field, while PAGD converges in the dual field. More details about the differences between these two types of approaches can be found in \cite{boyd2004convex}.


%
%


\begin{figure}[t]
		\centering
		\includegraphics[width=0.8\linewidth]{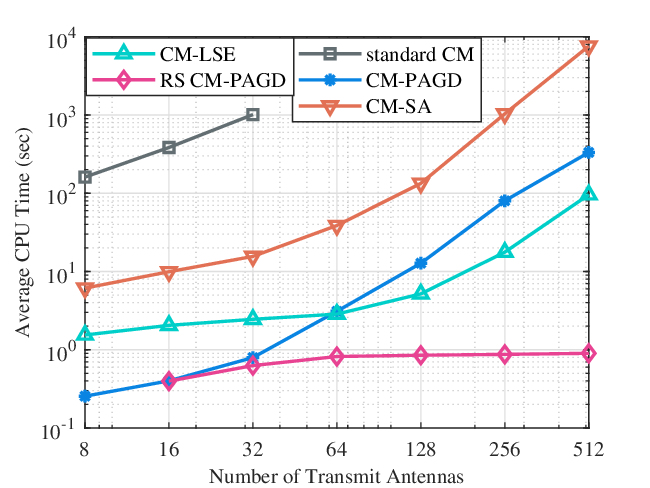}
	\caption{Averaged CPU time versus the number of transmit antennas when $G=3, K_G=4  $ and SNR$=20  $dB.}
	\label{img:performance_vs_Nt}
\end{figure} 
\begin{table}
	\begin{center}
		\caption{Weighted Sum Rate (bps/Hz) versus the Number of Transmit Antennas Comparison among Different Strategies.}
		\label{tab2}
	\scalebox{0.9}{	\begin{tabular}{ccccccc}
			\toprule
			$ L $  &   16& 32 & 64 & 128&256&512\\
			\hline\hline
			standard CM&13.7879&16.6894 & N/A&N/A & N/A&N/A\\
			\hline
			CM-PAGD &13.4855&16.4728 &18.8373&20.8721 & 22.9418&24.9572\\
			\hline
			CM-SA   &13.3432&16.0404 & 18.4644&20.7130 & 22.8966&25.0266\\
			\hline 
			CM-LSE&13.4231 &16.5445 & 18.6940 &20.7340 & 22.8947&25.0205\\
			\hline
			RS CM-PAGD&13.7854& 16.6882 & 18.8342&20.7431 & 22.8898&25.0221\\
			\bottomrule
		\end{tabular}}
	\end{center}
\end{table}

\par Table.\,\ref{tab2} shows the WSRs achieved by different algorithms versus the number of transmit antennas with transmit SNR$=20 $dB, and  Fig.$ \,\ref{img:performance_vs_Nt} $ illustrates the corresponding average CPU time. Due to the exponentially increasing computational cost of the standard CM with number of transmit antennas, we exclusively present its results for scenarios with 16 and 32 transmit antennas. Table.\,\ref{tab2} shows that all algorithms demonstrate nearly identical WSR performance across varying numbers of transmit antennas. Notably, the proposed RS CM-PAGD attains such excellent WSR performance while significantly reducing computational time. Its computational time is not proportional to the number of transmit antennas. This substantial decrease in computational complexity for XL-MIMO transceiver design marks the proposed X CM-PAGD as a promising algorithm for 6G.

\subsection{Comparison for Asymptotic Optimal Algorithms}
\begin{figure}[t]
	\centering
	\includegraphics[width=0.8\linewidth]{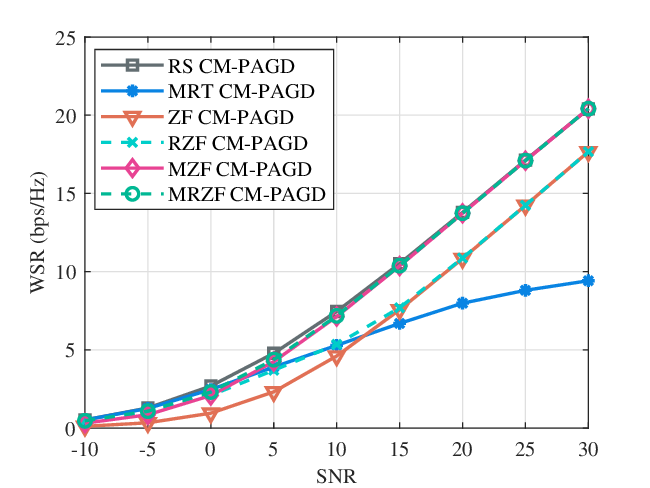}
	\caption{WSR versus the transmit SNR, when $ L=16 $, $ G=3 $ and $ K_G=4 $.}
	\label{img:sum_rate_vs_Nt}
\end{figure}  
\par Fig.$ \,\ref{img:sum_rate_vs_Nt} $ shows the WSR versus SNR comparison among different X CM-PAGD algorithms (X=MRT/ZF/RZF/MZF/MRZF) and RS CM-PAGD when the number of transmit antennas is $ L=16 $. It is evident that MRT achieves near optimal performance in the low SNR regime (i.e., SNR from $ -10 $ dB to $ 0 $ dB), while MZF and MRZF achieve asymptotically optimal performance in the high SNR regime (i.e., SNR is larger than $ 20 $ dB). The numerical results align with the theoretical analysis in Section \ref{sec:lowDimBF}. Also, it is observed that the classical ZF and RZF beamforming designs attain obvious performance loss compared to the near-optimal solution especially in the high SNR regime. This contrasts with the results in the unicast-only transmission \cite{Bjornson2014}, where ZF and RZF achieve near optimal performance. This is due to the distinct relations among user channels in the multi-group multicast communication, as discussed in Remark \ref{difference MZF and ZF}. 

\begin{figure}[t]
\vspace{-3mm}
	\centering	\includegraphics[width=0.8\linewidth]{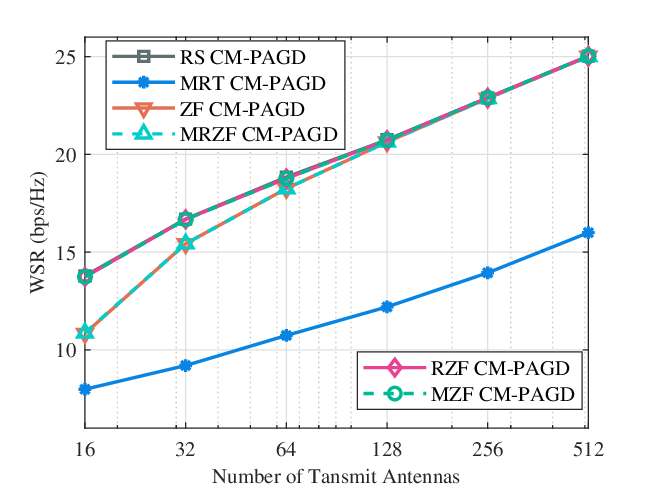}
	\caption{WSR versus the number transmit antennas, when  $ G=3 $, $ K_G=4 $ and SNR=20 dB.}
	\label{img:WSR_vs_antenna}
\end{figure}

Fig.$ \,\ref{img:WSR_vs_antenna} $ illustrates the WSR versus the number of transmit antennas when SNR$=20 $dB. As the number of transmit antennas increases exponentially, RS, ZF, RZF, and MRZF all achieve asymptotically optimal WSR performance. In contrast, MRT exhibits relatively poor WSR performance due to the asymptotic channel orthogonality. It suffers from severe performance degradation due to the presence of inter-group interference. 

\section{Conclusion}\label{Sec:concu}
\par In this study, we analyze the optimal and low-dimensional beamforming structures for a downlink multi-antenna multi-group multicast transmission network. Specifically, by leveraging the KKT conditions and SIT duality, we identify the optimal multi-group multicast beamforming structure for a general utility function-based maximization problem that embraces the WSR and MMF problems as special cases. This structure reveals valuable insights behind the multi-group multicast beamforming design, and inspires us to discover inherent low dimensional beamforming structures that are asymptotically optimal in various regimes of transmit SNR or the number of transmit antennas. These appealing beamforming structures provide guideline to efficient beamforming design for multi-group multicast transmission. Specially, we consider a special problem when the general utility function is the WSR. By exploiting the optimal beamforming structure, we propose an efficient optimization toolbox-free algorithm based on the CM framework and the proposed PAGD algorithm to solve the problem. We further exploit the low dimensional beamforming structures, and propose RS and MRT/ZF/RZF/MZF/MRZF based beamforming algorithms to further reduce the computational complexity. Numerical results demonstrate that these proposed algorithms achieve near-optimal performance while significantly reducing the computational complexity compared to the baseline schemes. They emerge as promising algorithms for ultra-massive MIMO applications in 6G. 

The developed algorithms are not limited to solving WSR problems, they can be easily extended to other utility functions-based optimization problems, such MMF and energy efficiency. Additionally, as the proposed PAGD algorithm effectively addresses the non-smooth property introduced in the multicast rate expressions, it can be easily extended to solve the problems of power domain non-orthogonal multiple access (PD-NOMA) and rate splitting multiple access (RSMA). This is due to the similarity between the rate expressions of the streams to be decoded by multiple users in PD-NOMA (and RSMA) and the multicast rate expressions from a mathematical perspective.
\bibliographystyle{IEEEtran}
\bibliography{efficient_MG_NOUM}
%
%

\end{document}